\documentclass[12pt]{article}
\usepackage{color}
\usepackage[latin1]{inputenc}
\usepackage[T1]{fontenc}
\usepackage[english]{babel}
\usepackage{amsfonts}
\usepackage[ruled]{algorithm2e}
\usepackage{amssymb}
\usepackage{textcomp}
\usepackage{afterpage}
\usepackage{mathtools}
\usepackage[allcolors=black]{hyperref}
\usepackage{threeparttable}
\usepackage{longtable}
\usepackage{array}
\usepackage{graphicx}
\usepackage{tabularx}
\usepackage{lipsum}
\usepackage{caption}
\usepackage{subcaption}
\usepackage{csquotes}
\usepackage{bbm}
\usepackage{dsfont}
\usepackage{booktabs}
\usepackage[title]{appendix}
\usepackage{epsfig}
\usepackage{lscape}
\usepackage[margin=1in]{geometry}
\usepackage{amsmath}
\usepackage{amsthm}
\usepackage{thmtools}
\usepackage{endnotes}
\usepackage{setspace}
\usepackage{float}
\usepackage{multirow}
\usepackage{comment}
\usepackage{chngpage} 
\usepackage{enumerate}

\usepackage{enumitem}
\DontPrintSemicolon

\def\P{\mathbb{P}}
\def\E{\mathbb{E}}
\DeclareMathOperator{\Var}{Var}
\DeclareMathOperator{\PML}{PML}

\DeclareMathOperator*{\length}{length}

\newtheorem{Theorem}{Theorem}[section]
\newtheorem{Definition}{Definition}[section]

\hypersetup{colorlinks=true,citecolor=blue,linkcolor=cyan}
\hypersetup{pdfstartview=FitH}
\hypersetup{pdfpagemode=UseNone}   
            
\usepackage[authordate,autocite=inline,backend=biber,maxcitenames=2, natbib,giveninits=true,uniquename=mininit]{biblatex-chicago}               
\DeclareFieldFormat[article]{citetitle}{#1}
\DeclareFieldFormat[article]{title}{#1} 

\makeatletter
\AtEveryBibitem{\global\undef\bbx@lasthash}
\makeatother
\providecommand{\keywords}[1]{\textbf{\textit{Keywords---}} #1}
\newcommand{\quotes}[1]{``#1''}
\usepackage{verbatim}
\makeatletter
\newcommand*{\rom}[1]{\expandafter\@slowromancap\romannumeral #1@}
\makeatother

\allowdisplaybreaks
\SetArgSty{textnormal}

\emergencystretch=\maxdimen
\usepackage{setspace}

\usepackage{longtable}
\usepackage{rotating}
\usepackage{pdflscape}

\newenvironment{longsidewaystable}[1]{\landscape\longtable{#1}}{\endlongtable\endlandscape}

\begin{document}

\thispagestyle{empty}
\author{\textsc{Roba Bairakdar}\footnote{Department of Mathematics and Actuarial Science, The American University in Cairo, Egypt
}
\and\textsc{Debbie J. Dupuis}\footnote{Department of Decision Sciences, HEC Montr\'eal, Canada
 }
\and \textsc{M\'elina Mailhot}
	\footnote{Department of Mathematics and Statistics, Concordia University, Canada
 }
 }
\title{\bf\LARGE Deviance Voronoi Residuals for Space-Time Point Process Models: An Application to Earthquake Insurance Risk} 
\date{}

\maketitle
\begin{sloppypar}
\setcounter{page}{1}

\begin{abstract}
Insurance risk arising from catastrophes such as earthquakes a component of the Minimum Capital Test for federally regulated property and casualty insurance companies. 
Analyzing earthquake insurance risk requires well-fitted spatio-temporal point process models. 
Given the spatial heterogeneity of earthquakes, the ability to assess whether the fits are adequate in certain locations is crucial in obtaining usable models. Accordingly, we extend the use of Voronoi residuals to calculate deviance Voronoi residuals. We also create a simulation-based approach, in which losses and insurance claim payments are calculated by relying on earthquake hazard maps of Canada. 
As an alternative to the current guidelines of OSFI, a formula to calculate the country-wide minimum capital test is proposed based on the correlation between the provinces. Finally, an interactive web application is provided which allows the user to simulate earthquake damage and the resulting financial losses and insurance claims, at a chosen epicenter location.
\end{abstract}

\newpage
\section{Introduction}
Catastrophic losses from earthquakes occurring in densely populated areas may cause a serious threat to the financial and economic stability of Property and Casualty (P\&C) insurance and reinsurance companies. Earthquakes happen randomly and without any reliable means of predicting their exact location or time of occurrence \autocite{usgspredict}. 
Unlike most meteorological disasters, earthquakes have very long return periods and accordingly small losses in recent years may not be indicative of future losses. 
Around half a million earthquakes are detected annually around the globe, with 
only 100 of them being strong enough to cause damage \autocite{cool_earthquake_facts_2021}. 
The infrastructure of buildings, their design and material are major factors that explain the extent of damage caused by an earthquake and its subsequent events. 

In Canada, earthquakes occur every year at a very high frequency, but on average, only one earthquake per week is large enough to be felt by residents \autocite{EQCanFAQ}. Earthquakes that are strong enough to cause material damage occur decades apart and they usually strike offshore or in locations that are not populated. 
Canada's west coast falls in the Circum-Pacific seismic belt, also known as the \quotes{the Ring of Fire}, where 81$\%$ of the world's largest earthquakes occur \autocite{usgsfaq}. 
On the other hand, 
seismic activity in Eastern Canada is attributed to regional stress fields and strong earthquakes occur at a relatively lower rate. 

According to the United States Geological Survey, earthquakes in Eastern North America can be felt at much further distances than earthquakes in Western North America of comparable size. 
Geologists attribute such differences to the nature of the underlying tectonic plates in the regions, and the size and age of buildings. Eastern North America has rocks that have been formed billions of years before the rocks in Western North America. These old rock formations are currently harder and denser due to their exposure to extreme pressures and temperatures and the faults have had more time to heal. Hence, seismic waves travel longer, compared to the younger faults in the west which absorb a lot of the seismic wave energy and minimize its spread \autocite{EastWestUSGS}.

\begin{figure}[!ht]
    \centering
    \includegraphics[width=\textwidth]{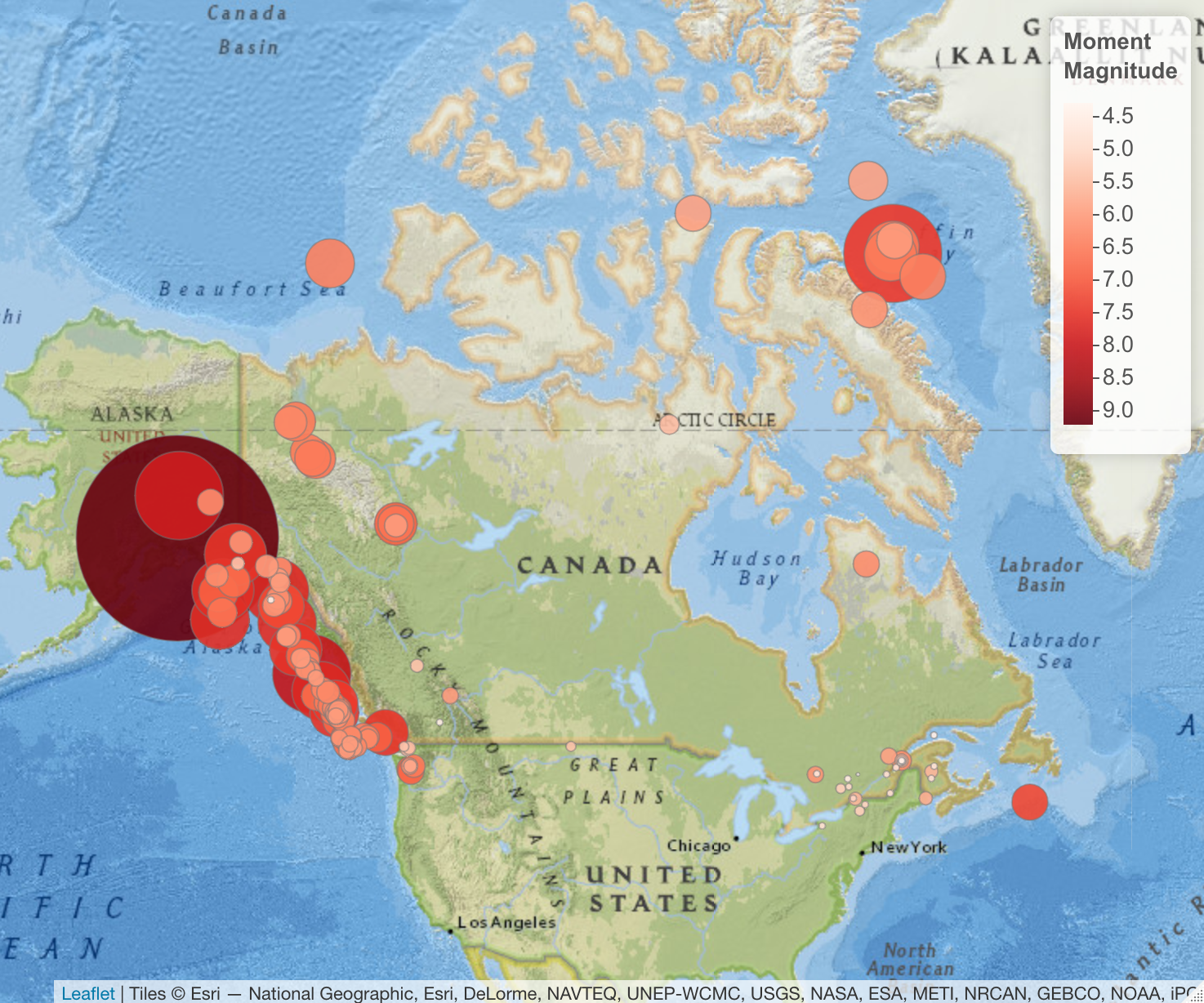}
    \caption{Significant Canadian earthquakes for the period $1600-2017$. The size and color of the circles are proportionate to the moment magnitude.}
    \label{Figure: Significant_EQ}
\end{figure}

Figure \ref{Figure: Significant_EQ} provides a spatial representation of the historical seismicity of significant earthquakes in Canada, for a total of 172 events for the period $1600-2017$ \autocite{lamontagne_halchuk_cassidy_rogers_2018}. An earthquake is considered significant if its moment magnitude exceeded 6, and/or it had been reported as felt by residents at Modified Mercalli intensity (MMI) of V or higher. The definitions of the MMI levels are available in Appendix \ref{Appendix: Tables MMI}. 
Given the actuarial nature of our study, we are interested only in significant earthquakes because claims are more likely to occur due to them. The magnitude of an earthquake is a value that describes its size, whereas the intensity is a measure of shaking, which varies across locations, depending mostly on nearness to the epicenter. The MMI level assigned to a location following an earthquake provides a more interpretable measure to non-seismologists than the magnitude because it quantifies the felt shaking. Initially, the MMI was a subjective measure approximated from humans' reports of felt ground shaking, but now, instrumental intensity measures are used to estimate the MMI, such as the ground motion-intensity relationship derived by \citet{wald1999relationships}.

Earthquake losses are not covered by a standard home insurance policy. Earthquake insurance usually applies a large deductible, 
and accordingly, the loss amount has to be substantially high before the insurance payments can be made, which may discourage homeowners from purchasing earthquake insurance coverage. British Columbia (BC) and the Ontario (ON) and Qu\'ebec (QC) region are at risk due to their large population density and elevated level of seismic activity. More than 70\% of the population in QC resides in seismic regions, yet only 3.4\% of homeowners in the cities of Montr\'eal and Qu\'ebec City hold earthquake insurance versus 65\% in the cities of Vancouver and Victoria \autocite{SwissReEastCanada}. 

The Office of the Superintendent of Financial Institutions (OSFI) is the Canadian organisation which sets guidelines for the recommended assessment of insurers' catastrophe risk management and calculation of probable maximum loss ($\PML$), the threshold dollar value of earthquake losses beyond which losses are unlikely. The insurance risk arising from earthquakes is one of the components of the Minimum Capital Test (MCT) for federally regulated P\&C insurance companies. 
The formula for the calculation of the earthquake reserve as set by \citet{OSFI} is
\begin{equation}\label{Eq: OSFIPML}
   \text{Country-wide} \PML_{1/500} = \left(\text{East Canada} \PML_{1/500}^{1.5} + \text{West Canada} \PML_{1/500}^{1.5}\right)^{\frac{1}{1.5}},
\end{equation}
where the earthquake PML is the gross PML, which is the PML amount after deductibles but before any reinsurance deductions. $\PML_{1/500}$ is a 1-in-500 year event, representing the $99.8^{\text{th}}$ percentile of the distribution of the annual maxima.

Due to the complexity of modeling seismic activity, Canadian insurers are expected to use earthquake models to calculate their earthquake reserve. AIR prepared an earthquake model for the insurance industry that provides estimates of the damages and economic losses due to all earthquake-related events for Canada. 
Another Canadian earthquake model is HazCan \autocite{Canada_HAZUS}, the Canadian version of HazUS \autocite{federal2013multi}.
These models provide a comprehensive study of probabilistic loss scenarios by relying on an inventory of population and building information as well as ground motion equations, but they are usually complex, sold at a high price, not available for public use and difficult to replicate. 

Spatio-temporal modeling of earthquakes has witnessed considerable progress, especially in seismically active regions in North America. Such models include the Regional Earthquake Likelihood Models (RELM) project \citep{field2007overview} and the Collaboratory for the Study of Earthquake
Predictability (CSEP) \citep{jordan2006earthquake}, which focus on the known faults in California. To assess the fitted models in the RELM project, \citet{schorlemmer2007earthquake} develop several numerical tests, such as the Number-test (or N-test), the Likelihood-test (or L-test) and the Likelihood-ratio-test (R-test). The N-test compares the total estimated and observed count of events, the L-test assesses the overall likelihood of the fitted model, and the R-test compares the relative performance of two fitted models. These methods are successful in assessing the overall fit of the model, however, they are incapable of identifying locations or time periods where the fit is poor. Recent progress in measuring the goodness-of-fit and comparing spatio-temporal models include pixel-based residual methods, in which the spatial region is divided into a predetermined regular grid and a residual is computed for each pixel, see \citet{baddeley2005residual, zhuang2006second}. Pixel-based residual analysis, such as raw residuals and Pearson residuals, show directly where the fitted models may be improved, while deviance residuals rank the performance of two fitted models. For models where the expected number of events at a pixel is close to zero, some problems may arise because of the heavy skewness of the distribution of the residual for the given pixel. To overcome these deficiencies, \citet{bray2014voronoi} use Voronoi polygons, generated by the observed spatio-temporal point pattern, which are adaptive to the inhomogeneity of the process, to compare the expected and observed count of events.

In this research project, we extend the Voronoi residual methods by calculating the Pearson Voronoi residuals, in analogy with Pearson residuals for generalized linear models. To compare competing models, we extend the definition of pixel-based deviance residuals to Voronoi polygons. In analogy with linear models, the resulting residuals may be called deviance Voronoi residuals. We use them to assess the goodness-of-fit of fitted spatio-temporal models for the significant Canadian earthquakes point pattern. Having established a well-fitting model, we analyze earthquake insurance risk by creating an open-source and reproducible simulation-based approach. An interactive web application allowing the user to simulate a significant earthquake based on any chosen location is also provided. Additionally, we review OSFI's MCT formula in eq.\ \eqref{Eq: OSFIPML} and provide a possible alternative.

The remainder of the paper is organized as follows. Section \ref{Section: STPP} provides a brief summary of spatio-temporal point processes, some of their properties and goodness-of-fit tests to assess the adequacy of fitted spatio-temporal models. Section \ref{Section: Methodology} explains the methodology used to calculate the building and content exposure, simulate a location and year of occurrence of an earthquake and estimate its financial impact. It also proposes an alternative approach to calculate the MCT for earthquake risk in Canada. Section \ref{Section: Results} compares our results with other results from simulated earthquakes in the literature, summarizes the outcome of our methodology and compares OSFI's MCT approach for earthquake risk to our alternative proposal. Section \ref{Section: Discussion} concludes the article. The unit of currency throughout is the Canadian dollar, unless otherwise stated.

\section{STPP and their Residuals}\label{Section: STPP}
A Spatio-Temporal Point Process (STPP) is a stochastic process that models data of the location and time of occurrence of events. Examples of processes that can be modeled include spread of diseases and pandemics, natural disasters such as earthquakes, tsunamis and volcanic eruptions. Point processes are studied thoroughly in time; see \citet{cox1980point, daley2003introduction} and in space; see \citet{cressiec, moller2003statistical, diggle2013statistical}. STPP are extensively used in seismology, but mainly to study earthquake aftershocks; see \citet{ogata1998space, zhuang2002stochastic}.

A STPP is defined as a random measure on a region $S\subseteq \mathbb{R}^2\times\mathbb{R}^{+}$ of space-time, where the spatial component is located in two spatial coordinates: longitude and latitude. A realization of a STPP is a spatio-temporal point pattern (stpp) consisting of location $\boldsymbol{x}_i$ (longitude and latitude) and a corresponding time of occurrence $t_i$ for an event, such that $\lbrace (\boldsymbol{x}_i,t_i): i =1, \ldots, n \rbrace$, where $(\boldsymbol{x}_i,t_i) \in A \times T$ for some known spatial region $A$ and temporal period $T$. 

The first-order properties of a STPP are defined by its spatio-temporal intensity function given by
\begin{equation}\notag
\lambda(\boldsymbol{x},t)=\lim_{\lvert d\boldsymbol{x}\rvert,\lvert dt\rvert \to 0}\frac{\mathbb{E}\left[ N(d\boldsymbol{x},dt) \right]}{\lvert d\boldsymbol{x}\rvert\lvert dt\rvert},
\end{equation}
where $d\boldsymbol{x}$ represents a small spatial region around the location $x$ such that $\lvert d\boldsymbol{x}\rvert$ is its area, $\lvert dt\rvert$ represents a small time interval containing the time point $t$ such that $\lvert dt\rvert$ is its length and $N(d\boldsymbol{x},dt)$ represents the number of events in $d\boldsymbol{x}\times dt$. Thus, $\lambda(\boldsymbol x,t)$ represents the mean number of events per unit area per unit of time. For a homogeneous STPP, $\lambda(\boldsymbol{x},t)=\lambda$ for all $(\boldsymbol{x},t) \in A\times T$. In practice, the STPP is observed on a finite space-time region $A\times T$, such that the marginal spatial and temporal intensities can be defined as 
\begin{equation}\notag
\lambda_A(\boldsymbol{x})=\int_T\lambda(\boldsymbol{x},t) dt, \quad \text{and} \quad \lambda_X(t)=\int_A\lambda(\boldsymbol{x},t) d\boldsymbol{x},
\end{equation}
respectively.

The N-test and the L-test are used to assess the adequacy of the fitted models \citep{schorlemmer2007earthquake}. For a fixed number $s$ of simulated realizations from the fitted model, the N-test requires computation of the fraction of simulations where the number of simulated events is less than the number of observed events. If the fraction is close to 0 or 1, the model is rejected. Similarly, the L-test requires computation of the fraction of simulated log-likelihoods that are less than the log-likelihood of the observed spatio-temporal point pattern. If the fraction is close to 0, the model is rejected. The two methods only consider the overall fit of the model and do not identify locations or time periods where the fit is poor. 

\begin{figure}[!ht]
    \centering
    \includegraphics[width=15cm]{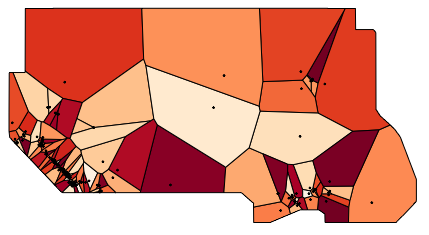}
    \caption{Voronoi tessellation of the significant Canadian earthquakes, shown in Figure \ref{Figure: Significant_EQ}.}
    \label{Figure: Voronoi_EQ}
\end{figure}

To overcome these deficiencies, \citet{baddeley2005residual} introduce pixel-based residual analysis methods for spatial point processes and \citet{zhuang2006second} extends to spatio-temporal point processes. Consider a STPP with conditional intensity $\hat\lambda(\boldsymbol x,t)$ at any location $\boldsymbol x$ and time of occurrence $t$. The simplest residual form, namely the raw residual, compares the total number of observed and estimated points within evenly spaced pixels over a regular grid, i.e.
\begin{equation}\notag
r^R(B_i) = N(B_i) - \int_{B_i}\hat\lambda(\boldsymbol x,t) dt d\boldsymbol x,
\end{equation}
where $N(B_i)$ is the number of points in bin $B_i$, which is a measurable set such that $B_i\subset A\times T$, for $i = 1,\ldots,m$, where $m$ is the total number of pre-determined bins. Thus, we can evaluate locations and time periods where the fitted intensity function fits poorly compared to the observed spatio-temporal point pattern. The residuals within each pixel can be standardized in numerous ways, which is important because otherwise residual plots can exaggerate the deviation in the pixels where the raw residual value is large. For example, the Pearson residuals are re-scaled raw residuals that have mean 0 and variance approximately equal to 1, in analogy with Pearson residuals in linear models, i.e.
\begin{equation}\notag
r^P(B_i) = \sum_{(t_j,\boldsymbol x_j)\in B_i} \dfrac{1}{\sqrt{\hat\lambda(\boldsymbol x_j,t_j)}} - \int_{B_i}\sqrt{\hat\lambda(\boldsymbol x,t)}dt d\boldsymbol x,
\end{equation}
for all $\hat \lambda(\boldsymbol x_i,t_i)>0$. See \citet{baddeley2005residual} for other methods of scaling the gridded residuals and \citet{baddeley2008properties} for their properties. Analogous to deviances for regression models, \citet{wong2009mainshock} propose using deviance residuals to differentiate model fits. Given two fitted intensity functions, $\hat\lambda_1$ and $\hat\lambda_2$, the deviance residual for pixel $B_i$ is the difference between their log-likelihoods, i.e.
\begin{align}
            r^D(B_i) &= \left(\sum_{(t_j,\boldsymbol x_j)\in B_i} \log(\hat\lambda_1(\boldsymbol x_j,t_j)) - \int_{B_i} \hat\lambda_1(\boldsymbol x,t) dt d\boldsymbol x  \right) \notag\\
            &- \left(\sum_{(t_j,\boldsymbol x_j)\in B_i} \log(\hat\lambda_2(\boldsymbol x_j,t_j)) - \int_{B_i} \hat\lambda_2(\boldsymbol x,t) dt d\boldsymbol x \right).\notag
\end{align}
Despite the mentioned advantages of pixel-based residuals, they can be problematic when the expected number of points at a pixel, based on a fitted model, is very small, causing heavy skewness of the distribution of its residual. This is problematic for earthquake occurrences modeling because the expected counts of events that are far from previous seismicity are usually zero. A possible solution could be to increase the size of the pixels such that the expected number of points in each pixel is large enough. However, this can be problematic for inhomogeneous processes because it is difficult, or almost impossible, to identify a one-size-fits-all pixel size. Accordingly, \citet{bray2014voronoi} propose Voronoi tessellations, which are adaptive to the inhomogeneity of the process. Consider an observed spatio-temporal point pattern consisting of location $\boldsymbol x_i$ and time $t_i$ for an event, such that $\lbrace (\boldsymbol{x}_i,t_i): i =1, \ldots, n  \rbrace$. A Voronoi tessellation is a partitioning of the spatial plane of interest into $n$ convex polygons, known as Voronoi polygons. Each polygon $C_i$ consists of the region of all locations on the spatial grid that are closer to $\boldsymbol x_i$ than to any other point in the process, for all $i=1, 2, \ldots, n$. Figure \ref{Figure: Voronoi_EQ} represents the Voronoi tessellations from Figure \ref{Figure: Significant_EQ}. For more details on Voronoi tessellations and their properties, see \citet{boots1999spatial}.
Given a fitted intensity model, $\widehat \lambda(\boldsymbol{x},t)$, raw Voronoi residuals are calculated by evaluating the raw residuals over the Voronoi polygons instead of the pixels. By definition, each Voronoi polygon contains exactly one observed point, i.e. $N(C_i)=1, \forall i=1, 2, \ldots, n$. Accordingly, the raw Voronoi residuals are defined by
\begin{align}
r^R_V(C_i) &= N(C_i) - \int_{C_i}\hat\lambda(\boldsymbol x,t) dt d\boldsymbol x, \notag\\
&= 1 - \int_{C_i}\hat\lambda(\boldsymbol x,t) dt d\boldsymbol x.\notag
\end{align}

Voronoi residual plots are constructed by shading each polygon according to the residual value under the distribution function of the modified gamma distribution, which is the approximate distribution of the raw residuals \citep{bray2014voronoi}. \citet{gordon2015voronoi} use the Voronoi raw residuals of a homogeneous Poisson process model as a benchmark or scale to assess the Voronoi raw residuals plots of alternative models.

In this research project, we propose to scale the raw Voronoi residuals, and accordingly calculate the Pearson Voronoi residuals, defined by
\begin{align}
   r^P_V(C_i) = \sum_{(t_j,\boldsymbol x_j)\in C_i} \dfrac{1}{\sqrt{\hat\lambda(\boldsymbol x_j,t_j)}} - \int_{C_i}\sqrt{\hat\lambda(\boldsymbol x,t)}dt d\boldsymbol x.\notag 
\end{align}
The fitted intensity must satisfy $\hat \lambda(\boldsymbol x_i,t_i)>0$ for all $\boldsymbol x_i$ and $t_i$ in order that the Pearson Voronoi residuals be well-defined. Accordingly, this will yield residuals that have mean 0 and variance approximately equal to 1. 

To compare competing models, we propose to extend the definition of pixel-based deviance residuals to Voronoi Polygons, i.e. to plot, in each Voronoi polygon, the difference between the log-likelihood for the two models. In analogy with linear models, the resulting residuals may be called deviance Voronoi residuals, defined by
    \begin{align}
            r^D_V(C_i) &= \left( \log(\hat\lambda_1(\boldsymbol x_i,t_i)) - \int_{C_i} \hat\lambda_1(\boldsymbol x,t) dt d\boldsymbol x  \right)- \left( \log(\hat\lambda_2(\boldsymbol x_i,t_i)) - \int_{C_i} \hat\lambda_2(\boldsymbol x,t) dt d\boldsymbol x \right) \label{Eq: DevianceVoronoiResiduals}
        \end{align}
Accordingly, we can see, for each polygon, where one model outperforms the other. Polygons with a positive (negative) deviance Voronoi residual indicate that $\hat \lambda_1$ ( $\hat \lambda_2$) provides a better fit. Summing the deviance residuals, i.e. $\sum_i r^D_V(C_i)$ yields a log-likelihood ratio score. This score provides a general indication of the improvement in fit provided by the best fitting model. Should one have more than two models, they can be compared two at a time to rule out the worst models. In Section \ref{Section: Methodology}, we apply the proposed methods to examine and compare several fitted spatio-temporal models to the significant Canadian earthquakes spatio-temporal point pattern.

\section{Earthquake Simulation Methodology} \label{Section: Methodology}
This section explains the methodology to calculate the building exposure and to simulate earthquake financial losses. We also propose an alternative formula to eq.\ \eqref{Eq: OSFIPML} for the MCT for earthquake risk in P\&C insurance companies.

\subsection{Calculating exposure and simulating earthquake damage} \label{Section: Procedure}

To estimate the seismic risk for a study area, we proceed with the following steps: 
\begin{enumerate}[label=(\roman*)]
\item Collecting building inventory and calculating exposure,
\item Simulating earthquakes in space and time,
\item Estimating ground shaking intensity, 
\item Calculating damage rates, and 
\item Estimating seismic loss and insurance claim payments. 
\end{enumerate}
Parts (ii) - (v) of the methodology are summarized in Algorithms \ref{Algorithm: EQSimulation} and \ref{Algorithm: EQLossesClaims}. To improve precision, the resolution of the analysis is at the Census Subdivisions (CSD) level, the municipalities used for statistical reporting purposes at Statistics Canada \autocite{StatCan_BoundaryFiles}. There are 5,162 CSD in Canada.\\

\textbf{\textit{(i) Collecting building inventory and calculating exposure}} \label{Section: Exposure}

For residential buildings, we calculate the total square footage by using the number of buildings in each residential building classification from the 2016 Canadian Census \autocite{statistics_Canada_2016} and the average square footage for each residential building from the Canadian Housing Statistics Program \autocite{statistics_Canada_2020_housing}. For each building type, the building exposure is calculated by multiplying the total square footage by the building replacement cost, in dollar units. The assumptions for replacement costs are taken from HazCan \autocite{Canada_HAZUS}.
The building construction price index (BCPI) provided by \citet{Statscan_BCPI} is used to inflate the construction costs to June 2021. To account for regional and embedded arbitrariness between construction companies, a random noise of $\pm 10\%$ is factored in the replacement cost. The building content replacement value for residential dwellings is assumed to be at $50\%$ of the building replacement cost \autocite{Canada_HAZUS, federal2013multi}. The building exposure based on the building types is then distributed based on the building construction material (wood, concrete, steel, masonry, etc.) by using HAZUS' general building scheme mapping information provided in Table 5.1 in \citet{federal2013multi}. 

Calculating the exposure of non-residential dwellings is not as straightforward because Statistics Canada currently does not have a comprehensive dataset for non-residential buildings. Accordingly, we rely on building permits data from \citet{Statscan_BuildingPermits} and use the annual ratios of institutional and governmental, commercial and industrial building permits to residential building permits. Hence, the non-residential buildings exposure is calculated as a percentage of the residential buildings total. Table \ref{Table:NonResiBuildings} in Appendix \ref{Section: Appendix Exposure} provides the suggested conversion of non-residential building types to HAZUS occupancy codes. As with residential buildings, the exposure is classified based on the construction material (wood, concrete, steel, masonry, etc.) by using HAZUS' general building scheme mapping information, provided in Table 5.1 in \citet{federal2013multi}. The building content replacement value for non-residential dwellings is assumed to be a percentage of the building replacement cost as suggested in \citet{federal2013multi}. 

\begin{figure}[!ht]
    \centering
    \includegraphics[width=15cm]{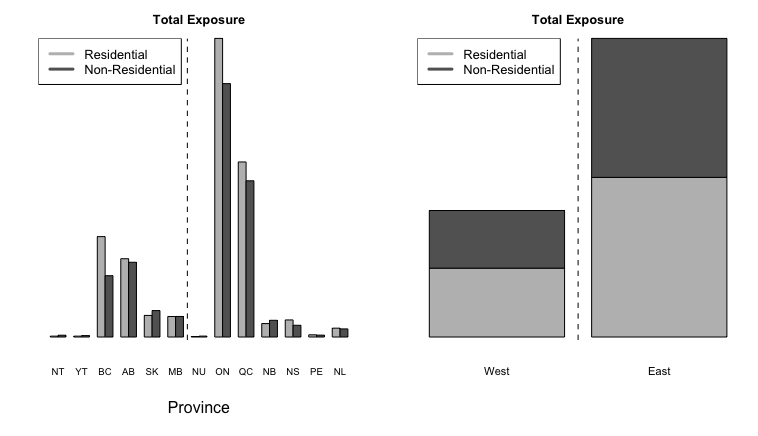}
    \caption{Total exposure per province (left) and for Eastern and Western Canada (right). The vertical dashed line splits Eastern and Western provinces.}
    \label{Figure: ExposureComparison}
\end{figure}

Figure \ref{Figure: ExposureComparison} summarizes the computed residential and non-residential exposure values per province. The values on the y-axis are omitted to maintain the confidentiality of CatIQ's (Canada's Loss and Exposure Indices Provider) exposure data, see Section \ref{Section: ExposureComparison} for details. Western Canada includes British Columbia (BC), Alberta (AB), Saskatchewan (SK), Manitoba (MB), Northwest Territories (NT) and Yukon (YT), while Eastern Canada includes Newfoundland and Labrador (NL), Nova Scotia (NS), Prince Edward Island (PE), New Brunswick (NB), Qu\'ebec (QC), Ontario (ON) and Nunavut (NU).

\begin{figure}[!ht]
    \centering
    \includegraphics[width=\textwidth]{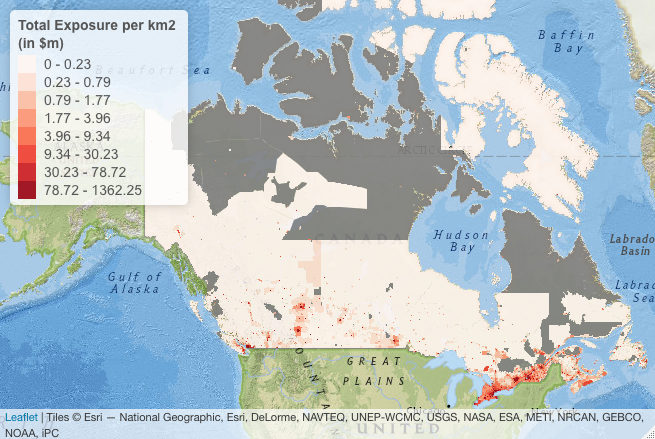}
    \caption{The calculated total exposure (building exposure and building contents exposure) of residential and non-residential buildings per km$^2$ for each CSD in Canada.}
    \label{Figure: ExposurePerKm2}
\end{figure}

Figure \ref{Figure: ExposurePerKm2} provides a spatial representation of our calculated total exposure, including residential and non-residential building exposure and building contents exposure, per km$^2$ for each CSD.\\

\textbf{\textit{(ii) Simulating earthquakes in space and time}}

By using the geographical coordinates and frequency of occurrence of the significant earthquakes dataset gathered by \citet{lamontagne_halchuk_cassidy_rogers_2018}, a spatio-temporal point pattern dataset is created. We choose earthquakes that occurred after the year 1900 because older data are not complete and hence can highly affect the temporal component of the STPP. There are 137 significant earthquakes for the period between 1900-2017. We assume that the space on which the events are realized is continuous and the temporal component is discrete because earthquake events are identified annually. We follow a non-parametric approach to estimate the spatio-temporal intensity function, $\hat\lambda(\boldsymbol x, t)$, by using kernel functions, as suggested by \citet{diggle1985kernel}. We also follow a practical approach and assume that the STPP is separable such that $\lambda(\boldsymbol x, t)$ can be factorized following
\begin{equation}
   \lambda(\boldsymbol x, t)=\lambda_{\boldsymbol X}(\boldsymbol x)\lambda_{T}(t), \label{Eq: Separatbility}
\end{equation}
for all $(\boldsymbol x,t) \in A\times T.$ Given our interest in the $\PML_{1/500}$, see eq. \eqref{Eq: OSFIPML}, which is a 1-in-500 year event, for earthquake damages, we are not interested in the exact time of occurrence of events, but rather the location of the epicenter and the affected communities. Hence, the separability assumption is valid for our project. Moreover, we will focus more on the spatial component, $\lambda_{\boldsymbol X}(\boldsymbol x)$ and perform the proposed Voronoi residual analysis methods on it. Spatial modeling and simulations are performed using the \texttt{splancs}, \texttt{spatstat}, \texttt{deldir}, and \texttt{stpp} packages in \textbf{\textsf{R}} \citep{splancs, spatstat, deldir, stpp, RCitation}, in addition to our functions for calculating the Voronoi residuals.

The temporal intensity function $\lambda_{T}(t)$ if estimated by using a Gaussian kernel with a bandwidth $h_T$ that follows Silverman's rule-of-thumb \citep{silverman1986density}, i.e.
$$h_T=0.9An^{-1/5},$$ where $A=\min\lbrace\text{sample standard deviation}, \text{sample interquartile range}/1.34\rbrace$.
As for the spatial intensity function $\lambda_{\boldsymbol X}(\boldsymbol x)$, we choose the quartic kernel \citep{diggle2013statistical}.
The choice of the spatial bandwidth $h$ determines the extent to which the intensity function will be smoothed, where larger bandwidths provide more smoothing. This choice involves a trade-off between bias and variance, such that larger bandwidths cause an increase in bias and a decrease in variance. The bias occurs because the estimator of the intensity function estimates a smoothed version of it rather than the intensity function itself. A large bandwidth requires a large number of points for estimation, thus reduces the estimation variance. The choice of a bandwidth is a controversial topic. We fit two different models for the spatial intensity function: both have the same kernel but with different bandwidths. The first bandwidth $h_1$ minimizes the estimated mean-square error of $\hat\lambda_{\boldsymbol X}(\boldsymbol x)$. The second bandwidth $h_2$ is based on cross-validation. Define the cross-validation likelihood by
\begin{equation}\notag
    LCV(h_2)=\sum_l\log(\lambda_{-l}(\boldsymbol x_l))-\int_A \lambda(\boldsymbol u) \text d \boldsymbol u, 
\end{equation}
where $\lambda_{-l}(\boldsymbol x_l)$ is the leave-one-out kernel-smoothing estimate of the intensity at $\boldsymbol x_l$ with smoothing bandwidth $h_2$, $\lambda(\boldsymbol u)$ is the kernel-smoothing estimate of the intensity at a spatial location $\boldsymbol u$ with smoothing bandwidth $h_2$ and $A$ is the spatial window of observation. \citet{baddeley2015spatial} observe that selecting a bandwidth by maximizing the LCV tends to choose more reasonable bandwidth values when the point pattern resembles tight clusters, such as our earthquake data. 

We assess the goodness-of-fit of the two fitted models, which we will call models $H_1$ and $H_2$, for the fitted quartic kernel with bandwidths $h_1$ and $h_2$, respectively. We also fit a homogeneous Poisson process, named model $P$, to the significant Canadian earthquakes spatial data set, where $\lambda(\boldsymbol{x})=\lambda$ for all $\boldsymbol{x} \in A$. Model $P$ will be used as a benchmark or scale to assess the Voronoi raw residuals plots of models $H_1$ and $H_2$.

\begin{figure}[!ht]
    \centering
    \includegraphics[width=\textwidth]{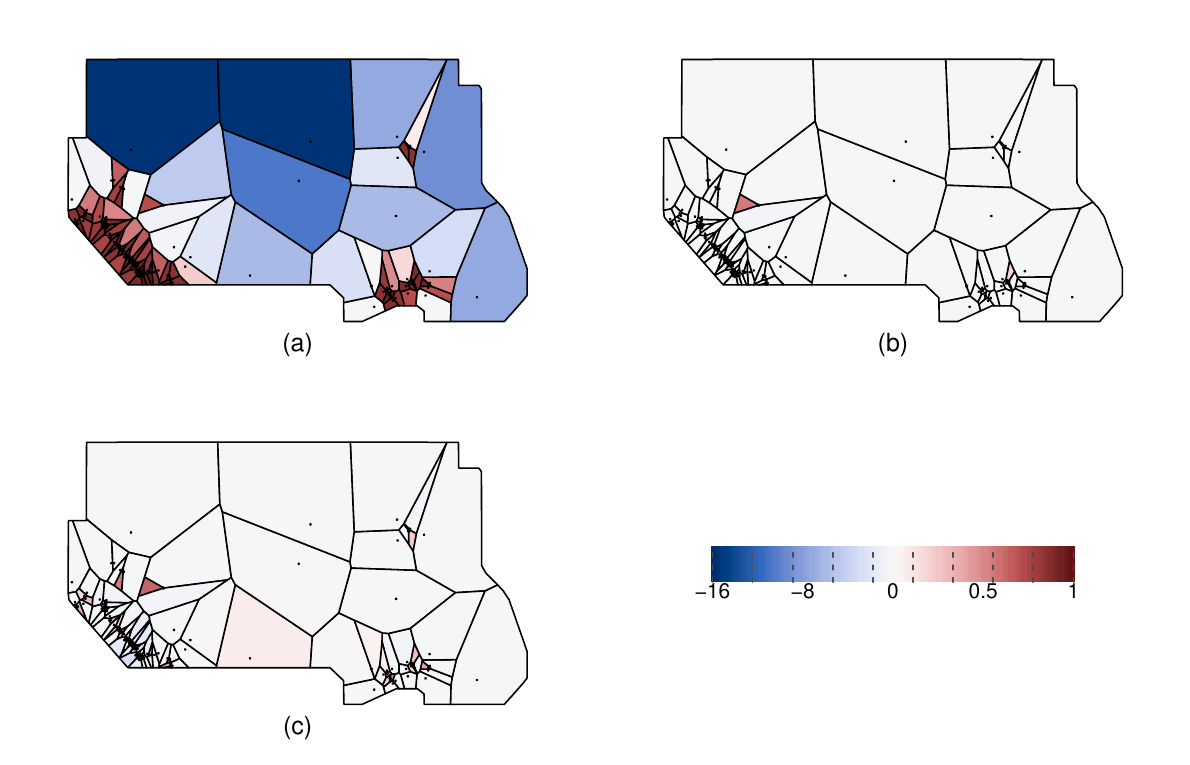}
    \caption{Raw Voronoi residuals of models $P$ (a), $H_1$ (b), and $H_2$ (c).}
    \label{Figure: Raw}
\end{figure}

Figure \ref{Figure: Raw} shows the raw Voronoi residual plots on the same color scale for all fitted models. An accepted model is one that has raw Voronoi residuals close to 0. By definition, the raw Voronoi residuals cannot exceed 1, however, they can be negative as the model can overestimate. Model $P$ is underestimating the expected count of earthquakes in the high-risk zones in Eastern and Western Canada, while it is overestimating the expected count in low-risk zones. We can expect a constant intensity function to perform poorly in such data sets because the events are clustered in the known faults and they are rare in other areas. Based on the color scale for models $H_1$ and $H_2$, both outperform model $P$, in fact, the raw residuals are close to 0 in almost all polygons. However, the better of the two models is unclear. Focusing on the high-risk zones, the fit of model $H_2$ is slightly poorer than model $H_1$, as shown in Figures \ref{Figure: RawWest} and \ref{Figure: RawEast} in Appendix \ref{Appendix: Section STPP}. A similar conclusion is obtained from the visual inspection of Pearson Voronoi residual plots, see Figures \ref{Figure: Pearson}, \ref{Figure: PearsonWest} and \ref{Figure: PearsonEast} in Appendix \ref{Appendix: Section STPP}. 

\begin{figure}[!ht]
    \centering
    \includegraphics[width=\textwidth]{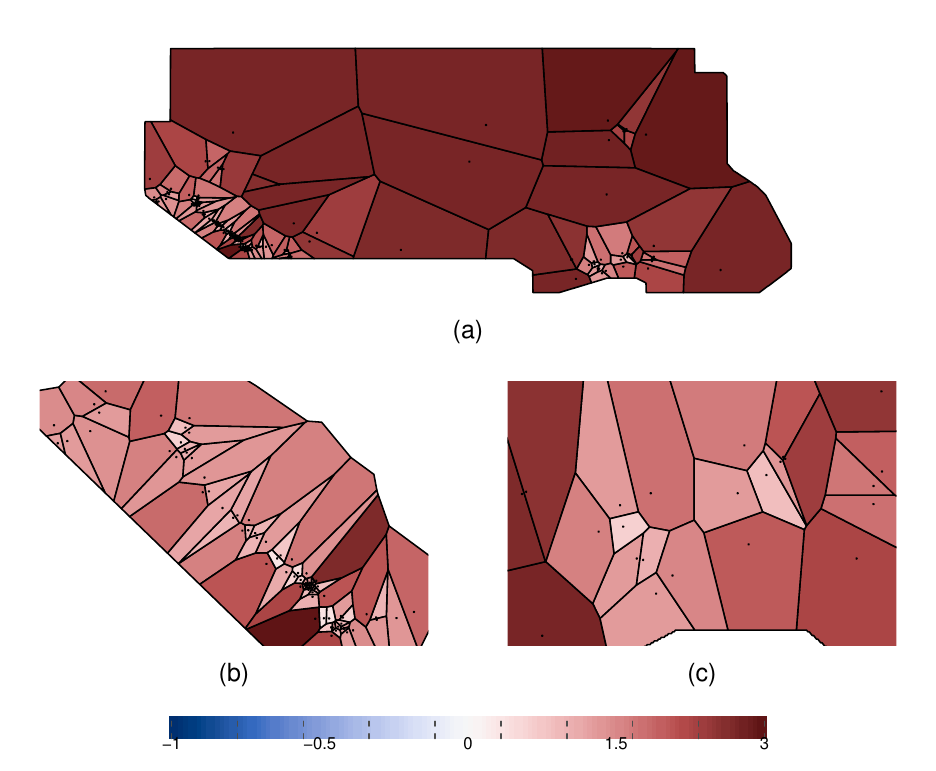}
    \caption{Deviance Voronoi residuals for model $H_1$ vs $H_2$ for Canada (a), Western Canada (b) and Eastern Canada (c).}
    \label{Figure: Deviance}
\end{figure}

To compare models $H_1$ and $H_2$, we use deviance Voronoi residuals, defined in eq. \eqref{Eq: DevianceVoronoiResiduals}. Figure \ref{Figure: Deviance} shows the deviance Voronoi residuals for model $H_1$ versus model $H_2$. Voronoi polygons with a positive (negative) deviance Voronoi residual indicate that model $H_1$ ($H_2$) provides a better fit. Inspecting the residual plot, we identify that model $H_1$ performs better than model $H_2$ in all polygons, and, in particular, model $H_1$ is performing much better than model $H_2$ in low-risk zones, such as Northern and Central Canada, where model $H_2$ tends to over-predict the frequency of occurrences. In both the Western and Eastern clusters of historical seismicity, more specifically, the Queen Charlotte fault (west) and the St. Lawrence paleo-rift faults (east), model $H_1$ is slightly outperforming model $H_2$. The polygon with the largest deviance residual is located in the Queen Charlotte fault, shown in Figure \ref{Figure: Deviance}(b). There are no regions where model $H_2$ fits better than model $H_1$, in fact, the smallest difference in the log-likelihood is 0.341, which is also located in the Queen Charlotte fault. Overall, the log-likelihood ratio score for Canada is 210.67, implying a large improvement for model $H_1$ compared to model $H_2$. The log-likelihood ratio scores for Canada for model $H_1$ versus $P$ is 586.56 and for model $H_2$ versus $P$ is 375.89, indicating that both $H_1$ and $H_2$ models provide considerably more accurate estimations.

The estimated spatial intensity function $\hat\lambda_{\boldsymbol X}(\boldsymbol x)$ for model $H_1$ and the temporal intensity $\hat\lambda_T(t)$ are multiplied to obtain the fitted spatio-temporal intensity function $\hat\lambda(\boldsymbol x, t)$, as given in eq. \eqref{Eq: Separatbility}. We follow the simulation algorithm in the \texttt{stpp} package in \textbf{\textsf{R}} to simulate a large number of years of significant earthquakes. For more details on the spatio-temporal simulation algorithm, see \citet{stpp}.\\

\textbf{\textit{(iii) Estimation of ground shaking intensity}}

\sloppy Earthquake hazard across Canada is evaluated by seismologists and geophysicists and the nation's seismic hazard maps are updated regularly. The National Building Code is subsequently revised to minimize the seismic risk on new infrastructures. Calculation of seismic risk requires the estimation of the ground shaking intensity, which can be defined by attenuation relationships of a given ground motion index, such as the peak ground acceleration (PGA), peak ground velocity, or spectral acceleration. The Geological Survey of Canada (GSC) provides seismic hazard values for a grid extending over Canada and surrounding areas for the 2015 National Building Code of Canada (NBCC) \autocite{PGAMap}. We use the mean PGA, which is available on firm soil sites, evaluated at the following probabilities of exceedance: $0.02, 0.01375, 0.0100, 0.00445, 0.0021, 0.0010, 0.0005$ and $0.000404$ per annum. For each grid point, a Generalized Pareto Distribution (GPD) is fitted such that the given quantiles represent the return levels that are exceeded every $50 - 2475$ years. The quantile of a GPD is shown in eq.\ \eqref{Eq: GPDQuantile} in Appendix \ref{Section: Appendix EVT}.

For a simulated earthquake location as in (ii), the GPD parameters of the grid point that is nearest to the earthquake epicenter are used to simulate a PGA value. To match the definition of significant earthquakes in Canada \autocite{lamontagne_halchuk_cassidy_rogers_2018}, the PGA value must have an associated moment magnitude M $>6$. The corresponding MMI level is calculated by using the relationship $\text{MMI} = 3.66 \log(\text{PGA}) -1.66$, with standard error
1.08, for PGA in cm/s$^2$ \autocite{wald1999relationships}. The corresponding moment magnitude M is calculated following Bakun's predicted distance attenuation equations,
\begin{itemize}
    \item Eastern Canada \autocite{BakunEast}:
    \begin{equation}\label{Eq: EastBakun}
    \text{M} = \frac{1}{1.68}\left(\text{MMI} - 1.41 + 0.00345\text{d} + 2.08\log_{10}(\text{d})\right),
    \end{equation}
    \item Western Canada \autocite{BakunWest}:
    \begin{equation}\label{Eq: WestBakun}
    \text{M} = \frac{1}{1.09}\left(\text{MMI} - 5.07 + 3.69\log_{10}(\text{d})\right),
    \end{equation}
\end{itemize}
where d is the distance in kilometers from the epicenter of the earthquake to its nearest neighbor on the PGA grid. Even though an earthquake has a single magnitude, it has multiple intensity values, depending on the location from the epicenter. Accordingly, Bakun's predicted distance attenuation equations are used to produce isoseismal maps, which are maps that identify areas of equally felt seismic intensity. These are estimated by calculating the distance from the epicenter for each MMI level, based on eq.\ \eqref{Eq: EastBakun} and eq.\ \eqref{Eq: WestBakun}, and MMI circles with radii equivalent to the calculated distances are graphed accordingly. The MMI levels are very important information that are required to quantify the percentage of damage caused by an earthquake. We consider the uncertainty in the estimation by applying random noise with the respective standard errors, obtained from \citet{BakunWest, BakunEast, wald1999relationships}. The CSDs inside each MMI circle are identified and the percentage of their affected areas are calculated for each seismic intensity level.\\

\textbf{\textit{(iv) Calculating damage rates}}

Earthquake damage can result in financial losses, such as the cost of damage repair and losses due to business interruptions, in addition to non-financial losses, such as fatalities and injuries. Each category of losses is divided further into direct losses from the damage caused by the ground shake and indirect losses from damage due to other hazards induced by the earthquake, such as a tsunami, landslide, liquefaction and/or a fire. This distinction is of interest to insurers to process claims depending on the coverage and policy conditions. In this article, only the direct financial losses that occur due to earthquake damage to buildings are considered. Building damage is split into structural damage and non-structural damage. Structural damage is the damage to the skeleton of the building, such as the roof and load-bearing walls. Non-structural damage can affect drift-sensitive components, such as non-bearing walls/partitions, veneer and finishes, acceleration-sensitive components, such as piping systems, elevators and lightning fixtures, and finally building contents, such as bookcases, office equipment and furnishings.

\citet{ATC1985} provides a benchmark study that relies on MMI to quantify the degree of ground shaking. Motion-damage relationships are developed, where the probability of being in a defined damage state for different levels of MMI is calculated. To quantify the damage due to an earthquake, MMI-based damage probability matrices (DPMs) for structural and non-structural damage for each building classification are required. These matrices provide for each building class the probability that a building is in one of seven damage states, given the MMI at the location of the building. The damage states and their corresponding damage factor range, which is the range of the percentage of exposure value that is damaged, are shown in the first two columns of Table \ref{Table: DPMExample}. \citet{thibert2008methodology} developed DPMs for structural and non-structural damage in BC by assuming that the buildings are nearly regular in shape, founded on firm ground, and are designed based on a National Building Code prior to 1990. Table \ref{Table: DPMExample} provides an example of a DPM.

\begin{table}[!ht]
  \begin{threeparttable}
    \centering\captionof{table}{DPM for Structural Damage in Wood Light Frame Residential building \autocite{thibert2008methodology}}\label{Table: DPMExample}
     \begin{tabular}{|l|>{\raggedleft\arraybackslash}p{4cm}|l|l|l|l|l|l|l|}
\hline
\textbf{Damage state}&\multicolumn{1}{>{\centering\arraybackslash}p{4cm}|}{\textbf{Damage factor range}}& \textbf{\rom{6}} & \textbf{\rom{7}} & \textbf{\rom{8}}& \textbf{\rom{9}}& \textbf{\rom{10}}& \textbf{\rom{11}}&\textbf{\rom{12}}\\
\hline
No damage &0 &0.08&0.04&0.01&***&***&***&*** \\
Slight damage&$0-1$  &0.75&0.28&0.06&0.01&***&***&*** \\
Light&$1-10$&0.17&0.64&0.86&0.69&0.19&0.02&***\\
Moderate&$10-30$ &***&0.04&0.05&0.20&0.76&0.69&0.42 \\
Heavy&$30-60$  &***&***&0.02&0.10&0.12&0.25&0.50 \\
Major&$60-100$ & ***&***&***&***&0.02&0.04&0.06 \\
Destroyed&$100$ &***&***&***&***&***&***&0.02 \\
\hline
\end{tabular}
    \begin{tablenotes}
      \small
      \item *** represents very small probability values, almost 0.
    \end{tablenotes}
  \end{threeparttable}
\end{table}

The percentage of damage for each building classification is defined in terms of the mean damage factor (MDF), which quantifies the expected damage as a percentage of exposure value. The MDF is the expected value of the damage given an MMI level, and is calculated by adding up the product of the damage factor by its corresponding probability. To embed further variability in the simulations, the damage factor is randomly drawn from each damage range. For each CSD in an MMI circle, the MDF for each damage type (structural (S), acceleration-sensitive (AS) non-structural, drift-sensitive (DS) non-structural, and building contents (BldgC)) is estimated for each building class.\\

\textbf{\textit{(v) Estimating seismic losses and insurance claim payments}}

The losses are calculated by summing the losses generated by S, DS, AS, and BldgC damages. Following \citet{onur2005regional}, we split the building exposure over damage types following: 25\% of the building exposure is for S, 37.5\% is for DS and 37.5\% is for AS components. Accordingly, the losses for the $j^\text{th}$ CSD are simply the product of the MDF for a given MMI level and the exposure of the CSD at that MMI level such that 
\begin{align*}
    \text{L}_{j,k,S} &=  \left[\text{MDF}_{S,k}\times \left(0.25\times\text{building exposure at CSD}_{j,k}\right) \right],\\
    \text{L}_{j,k,DS} &=  \left[\text{MDF}_{DS,k}\times \left(0.375\times\text{building exposure at CSD}_{j,k}\right) \right],\\
     \text{L}_{j,k,AS} &=  \left[\text{MDF}_{AS,k}\times \left(0.375\times\text{building exposure at CSD}_{j,k}\right) \right],\\
     \text{L}_{j,k,BldgC} &=  \left[\text{MDF}_{BldgC,k}\times\text{building contents exposure at CSD}_{j,k} \right],
\end{align*}
where $k$ is an index for the $k^\text{th}$ building class. Thus, 
\begin{equation}\notag
    \text{Total Losses}_{j,k} = \text{L}_{j,k,S} + \text{L}_{j,k,DS} + \text{L}_{j,k,AS} + \text{L}_{j,k,BldgC}.
\end{equation} Note that a CSD can be affected by multiple MMI levels, and accordingly the percentage area affected at each MMI level is used to distribute the exposure of a CSD.

The insurance terms used in this article are based on the deductibles, policy limits and insurance market penetration rates provided in \citet{worldwide2013study}, which are displayed in Table \ref{Table: InsuranceTerms}. AIR Worldwide have only given statistics for QC and BC, and hence, we assume the smallest market penetration for the provinces for which we have no information. Insurance claims for the $j^\text{th}$ CSD are calculated following
\begin{align*}
\text{ClaimPmt}_{j,k} &= \begin{cases}
0 , &\text{Total Losses}_{j,k} \leq d_{j,k},\\
\pi_j\left(\text{Total Losses}_{j,k} - d_{j,k} \right), &d_{j,k}<\text{Total Losses}_{j,k} \leq u_{j,k},\\
\pi_{j}\left(u_{j,k} - d_{j,k} \right), &\text{Total Losses}_{j,k} > u_{j,k},\\
\end{cases}
\end{align*}
where 
\begin{align*}
\pi_j &= \% \text{ earthquake insurance market penetration at CSD}_{j},\\
d_{j,k} &= \% \text{ deductible}_{j} \times \sum_k \left(\text{total exposure at CSD}_{j,k} \right),\\
u_{j,k} &= \% \text{ policy limit}_{j} \times \sum_k \left(\text{total exposure at CSD}_{j,k} \right),
\end{align*}
where $k$ is an index for the $k^\text{th}$ building class and total exposure for a CSD includes building content exposure. Finally,
\begin{align*}
   \text{Total Losses}_j &= \sum_k \text{Total Losses}_{j,k}, \\
   \text{ClaimPmt}_{j} &= \sum_k \text{ClaimPmt}_{j,k}. 
\end{align*}
Appendix \ref{Appendix: Algorithms} summarizes the methodology in two algorithms: Algorithm \ref{Algorithm: EQSimulation} includes the steps required to simulate an earthquake and Algorithm \ref{Algorithm: EQLossesClaims} explains how to calculate the losses and the insurance claims for each simulated earthquake.
 
\begin{table}[!ht]
  \begin{threeparttable}
\captionof{table}{Assumptions for the earthquake insurance terms and market penetration rates, as prescribed in  \citet{worldwide2013study}}\label{Table: InsuranceTerms}
     \begin{tabular}{|>{\raggedleft\arraybackslash}l|l|>{\raggedleft\arraybackslash}p{4.1cm}|>{\raggedleft\arraybackslash}p{2.4cm}|>{\raggedleft\arraybackslash}p{1.3cm}|}
\hline
\multicolumn{1}{|c|}{\textbf{Property type}} & \multicolumn{1}{c|}{\textbf{Location}}& \multicolumn{1}{>{\centering\arraybackslash}p{4.1cm}|}{\textbf{Market penetration rate }}  &\multicolumn{1}{c}{\textbf{Deductible}}& \multicolumn{1}{|c|}{\textbf{Limit}} \\
\hline
\multirow{6}{*}{{Residential}} & Vancouver Metro& 55\% & 10\% & 100\%\\
& Victoria Metro& 70\% & 8\% & 100\%\\
& Rest of BC & 40\%& 8\% & 100\%\\
& Montr\'eal Metro &5\%& 5\% & 100\%\\
& Qu\'ebec Metro &2\%& 5\% & 100\%\\
& Rest of QC &2\%& 5\% & 100\%\\
\hline
& Vancouver Metro & 85\%& 10\% & 80\%\\
& Victoria Metro & 85\%& 7.5\% & 80\%\\
{Commercial /} & Rest of BC & 85\%& 7.5\% & 80\%\\
{Industrial} & Montr\'eal Metro & 60\%& 5\% & 80\%\\
& Qu\'ebec Metro & 60\%& 5\% & 80\%\\
& Rest of QC & 60\%& 5\% & 80\%\\
\hline
\end{tabular}
    \begin{tablenotes}
      \small
      \item Market penetration rates are percentages of the total exposure value.
      \item Deductibles and policy limits are percentages of the insured exposure value.
    \end{tablenotes}
  \end{threeparttable}
\end{table}

\subsection{MCT for earthquake risk} \label{Section: Proposal}

OSFI's MCT formula in eq.\ \eqref{Eq: OSFIPML} is a function of the PML in Eastern and Western Canada. Actuaries define the PML as the worst case scenario of the losses. Let $X_1, \ldots, X_n$ be a sequence of independent random variables having a common distribution function $F$ and consider $M_n = \max\lbrace X_1,\ldots,X_n\rbrace$. There are several suggestions in the actuarial literature to calculate the PML, such as $(1+\theta)\E[M_n]$ or $\E[M_n]+\theta \sqrt{\Var[M_n]}$, as suggested by \citet{wilkinson1982estimating,kremer1990probable, kremer1994more}, where $\theta$ is a chosen constant. The PML is mathematically defined to be an extreme quantile of $M_n$ such that 
\begin{equation} \label{Eq: PMLCDF}
    \P(M_n \leq \PML_\epsilon) = 1-\epsilon
\end{equation}
for some small $\epsilon >0$. Accordingly, the PML is the threshold dollar value of losses beyond which losses are highly unlikely.

As explained in Appendix \ref{Section: Appendix EVT}, under some conditions, the distribution of the exceedances over a high threshold $u$ can be approximated by a GPD $G_{\xi,\sigma}$ and the number $N$ of exceedances over $u$ follows a Poisson distribution with rate $\lambda$. The distribution of the maximum of the $N$ exceedances can be approximated by a Generalized Extreme Value (GEV) distribution $H_{\xi,\mu,\psi}$ with $\mu = \frac{\sigma}{\xi}\left(\lambda^\xi -1\right)$ and $\psi = \sigma \lambda^\xi$ \autocite{cebrian2003}. Solving for the $\PML$ in eq.\ \eqref{Eq: PMLCDF} yields
\begin{equation} \label{Eq: PMLQuantile}
    \PML_\epsilon = u + \frac{\sigma}{\xi}\left[\left( -\frac{\lambda}{\ln(1-\epsilon)}\right)^{\xi} -1  \right].
\end{equation}
See Section \ref{Section: MaxofGPD} in Appendix \ref{Section: Appendix EVT} for details.

Inspired by \citet{SolvencyII}, which guides the European standards for capital requirements, we propose using the correlation of the insured losses between Canadian provinces and territories to calculate the minimum capital requirements for earthquake risk in Canada. The correlation coefficients should reflect the dependence between earthquake financial risks in Canadian provinces and territories. The Solvency \rom{2} capital requirements for earthquake insurance risk adapted to Canada is

 \begin{equation} \label{Eq: CorrPML}
     \text{Country-wide} \PML_{1/500} = \sqrt{\sum_{r,s}\text{CorrEQ}_{r,s}\times \PML_{1/500,r} \times \PML_{1/500,s}},
 \end{equation}
where: 
\begin{enumerate}
    \item the sum includes all possible combinations $(r,s)$ of Canadian provinces and territories.
    \item $\text{CorrEQ}_{r,s}$ denotes the correlation of the insured losses for earthquake risk for province $r$ and province $s$
    \item $ \PML_{1/500,r}$ and $ \PML_{1/500,s}$ denote the Gross $\PML$ for province $r$ and $s$ respectively, which is the $\PML$ amount after deductibles but before any reinsurance deductions.
\end{enumerate}

The $\PML_{1/500}$ is calculated both empirically from the simulations by computing the $\left(1-1/500\right)$ quantile of the annual maximum losses, which may be zero for years with no earthquake damage and  by substituting estimates of the parameters $\mu, \sigma, \xi,$ and $\lambda$ in eq.\ \eqref{Eq: PMLQuantile}. The correlation matrix $\text{CorrEQ}$ is created by calculating the pairwise correlation of the simulated financial losses. We should have high correlation when provinces are affected concurrently with the same earthquake. Several correlation measures were attempted: Pearson's correlation coefficient, Kendall's tau and Kendall's tau for zero-inflated continuous variables \autocite{pimentel2015association}. The latter was considered because of the nature of the data where a large number of observations have no financial losses because earthquakes do not affect all provinces in Canada simultaneously\footnote{As discussed in \citet{denuit2017bounds}, there are bounds on Kendall's tau for zero-inflated continuous variables such that the correlation values are not between $[-1,1]$, but rather have a smaller range. Accordingly smaller Canada-wide PML values will be obtained, compared to the other methods. Hence, this correlation method was ultimately deemed inappropriate for the purpose of this article.}.

\section{Results} \label{Section: Results}
In this section we compare our results from the proposed methodology in Section \ref{Section: Methodology} to other sources and articles. Section \ref{Section: ExposureComparison} compares the building exposure values following our methodology to the exposure values obtained from CatIQ, Canada's insured loss and exposure indices provider. Section \ref{Section: ComparisonOtherEQ} compares the financial losses and insurance claims generated from our algorithm to another earthquake study in the literature; see \citet{onur2005regional}. Finally, Section \ref{Section: SimulationResults} provides a detailed analysis of the earthquake simulation results. 

\subsection{Comparison of exposure values}\label{Section: ExposureComparison}

\begin{figure}[!ht]
    \centering
    \includegraphics[width=\textwidth]{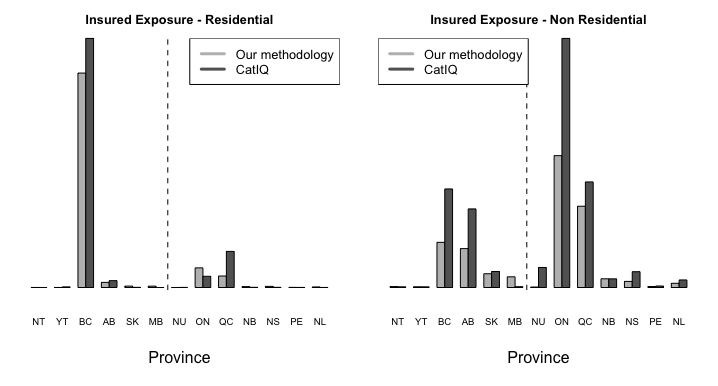}
    \caption{Calculated insured exposure, as explained in Section \ref{Section: Exposure}, vs the exposure collected by CatIQ.}
    \label{Figure: ExposureCatIQ}
\end{figure}

CatIQ collects values of exposure, insured losses and other related information to serve the needs of insurers, reinsurers and other stakeholders through an online subscription-based platform. Figure \ref{Figure: ExposureCatIQ} compares the exposure values obtained from CatIQ for 2020 to the insured exposure calculated by applying the earthquake insurance market penetration rates from Table \ref{Table: InsuranceTerms} to the cost of building replacement, as explained in Section \ref{Section: Exposure}. The comparison is made in June 2021 dollar value. The values on the y-axis are removed to maintain the confidentiality of CatIQ's data, which represents aggregated exposure values from a majority of Canadian insurers. Our methodology is based on estimations and several assumptions, while CatIQ's exposure values are an accurate representation of the insured exposure for the insurers who submitted their data. Given that we intend to provide an open-source earthquake insurance risk assessment tool that can be easily explained and there are no publicly available exposure data, the insured exposure values  for residential buildings calculated with our methodology are deemed sufficient for this paper. The non-residential exposure differs considerably, which is explained by the use of many assumptions to estimate the values; see Section \ref{Section: Exposure}. 

\subsection{Comparison of the simulated financial losses with other earthquake studies}\label{Section: ComparisonOtherEQ}

In this section, we compare the results of the methodology explained in Section \ref{Section: Methodology} to calculate direct financial losses and insurance claims with those estimated in earthquake studies in the literature.

Conducted in 2001 and published in 2005, the study by \citet{onur2005regional} estimates the financial losses for the City of Victoria and the City of Vancouver, subject to an earthquake of MMI \rom{8} affecting the entirety of the cities. Total monetary losses of \$430 million for the City of Victoria and \$3.5 billion for the City of Vancouver, in 2001 Canadian dollars, are estimated. Inflated to June 2021 by using the BCPI from \citet{Statscan_BCPI}, the values correspond to \$835 million for the City of Victoria and \$6.8 billion for the City of Vancouver. Our methodology, under the same MMI level assumption, produces losses of \$941 million for the City of Victoria and \$5.73 billion for the City of Vancouver. The differences can be attributed to two factors: the losses calculated by \citet{onur2005regional} excluded the losses resulting from damage to building contents, and different sources in calculating the construction costs where \citet{onur2005regional} used values provided by local construction companies.

\subsection{Simulation results} \label{Section: SimulationResults}
In this section, we give the results of our earthquake simulations. We also compare the value of the Canada-wide PML for earthquake risk in P\&C companies obtained from our proposed method in eq.\ \eqref{Eq: CorrPML} to that computed from OSFI's formula in eq.\ \eqref{Eq: OSFIPML}.

100,000 years of earthquake simulations are performed by relying on the procedure detailed in Section \ref{Section: Methodology}. The number of years is chosen to be high enough to minimize the variance in the results\footnote{The results in this section are compared against another set of 100,000 years and minimal variations were observed.}. Table \ref{Table: PercentageYearsEQ} summarizes the proportion of years that had a significant earthquake and the proportion of years that had a damage-inducing significant earthquake.

\begin{table}[!ht]
\begin{center}
\centering\captionof{table}{The proportion of years with a significant earthquake and the proportion of years with a significant earthquake causing damage, based on 100,000 simulated years of earthquakes.}\label{Table: PercentageYearsEQ}
\begin{tabular}{|>{\centering\arraybackslash}p{3.5cm}|>{\raggedleft\arraybackslash}p{4.1cm}|>{\raggedleft\arraybackslash}p{7.2cm}|}
\hline
\multicolumn{1}{|>{\centering\arraybackslash}p{3.5cm}|}{\textbf{Number of earthquakes ($n$) }}
&\multicolumn{1}{>{\centering\arraybackslash}p{4.1cm}|}{\textbf{\% of years with $n$ earthquakes  }} 
 & \multicolumn{1}{>{\centering\arraybackslash}p{7.2cm}|}{\textbf{\textbf{\% of years with $n$ earthquakes causing damage}}} \\
\hline
0&45.651& 59.536\\
1&27.127& 26.321\\
2&15.265& 9.886\\
3&7.127& 3.117\\
4&2.999& 0.846\\
5&1.177& 0.227\\
6&0.425& 0.057\\
7&0.160& 0.01\\
8&0.065& 0\\
9&0.013& 0\\
10&0.003& 0\\
11&0.001& 0\\
12+&0&0\\
\hline
\end{tabular}
\end{center}
\end{table}

\begin{figure}[!ht]
    \centering
    \includegraphics[width=\textwidth]{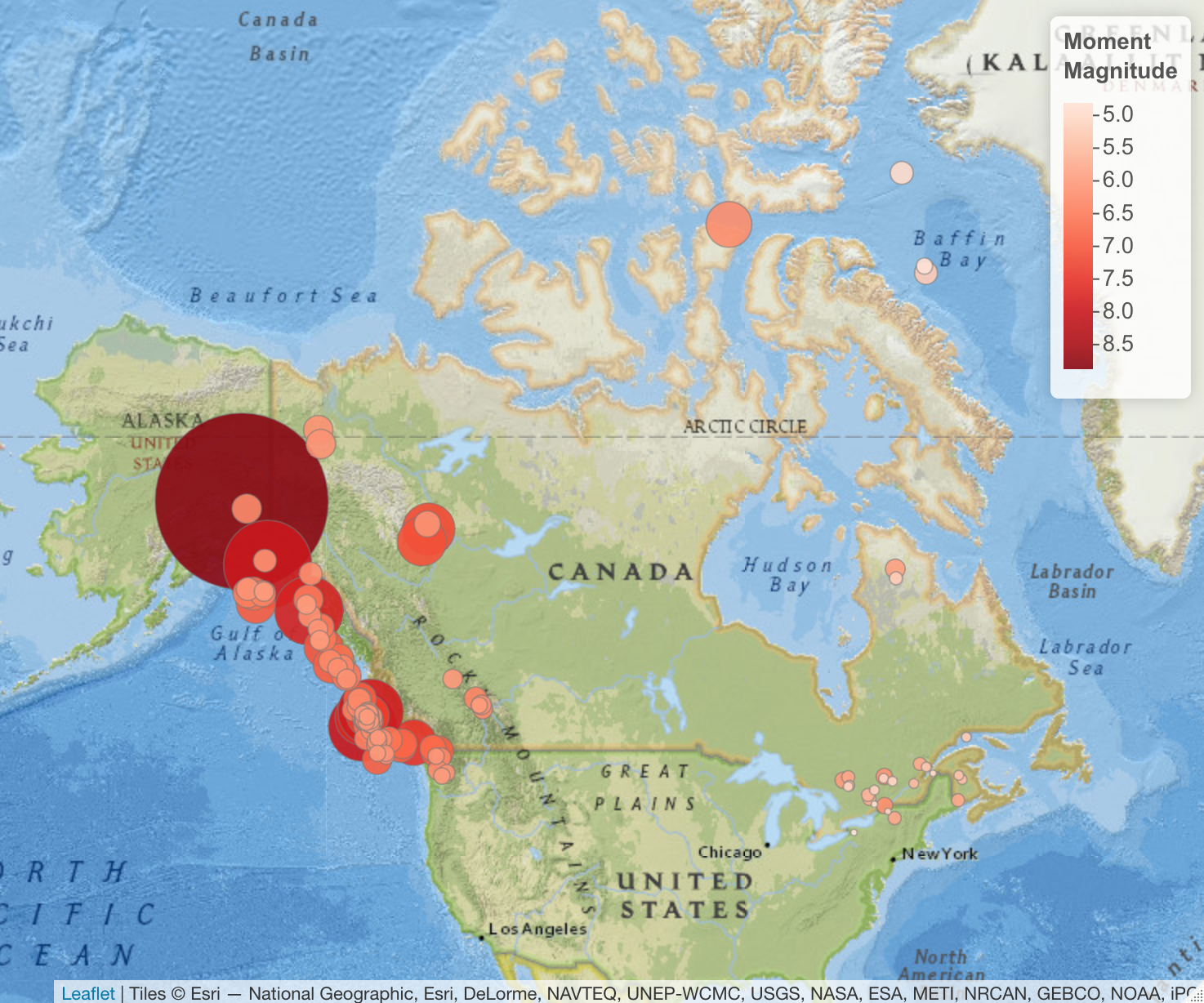}
    \caption{200 years of simulated earthquakes. The size and color of the circles are proportionate to the moment magnitude.}
    \label{Figure: Simulation200Years}
\end{figure}

Figure \ref{Figure: Simulation200Years} shows the locations and moment magnitudes of a sample of 200 years from the simulations. There is a great similarity with the historical seismicity shown in Figure \ref{Figure: Significant_EQ} in terms of the locations and intensity of the earthquakes.

\begin{figure}[!ht]
\centering
\begin{subfigure}{.5\textwidth}
  \centering
  \includegraphics[width=0.99\linewidth]{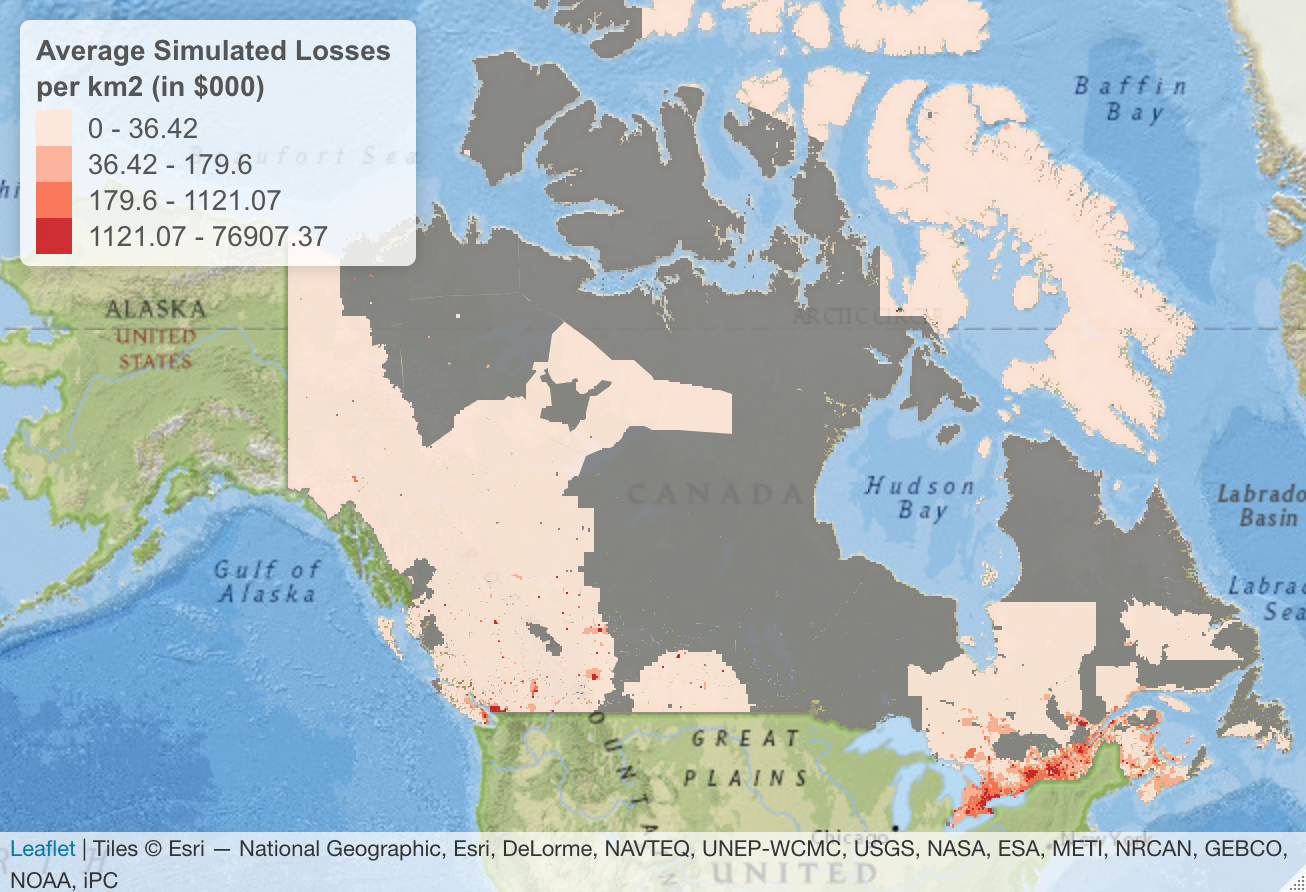}
  \caption{Average Simulated Losses}
\end{subfigure}%
\begin{subfigure}{.5\textwidth}
  \centering
  \includegraphics[width=0.99\linewidth]{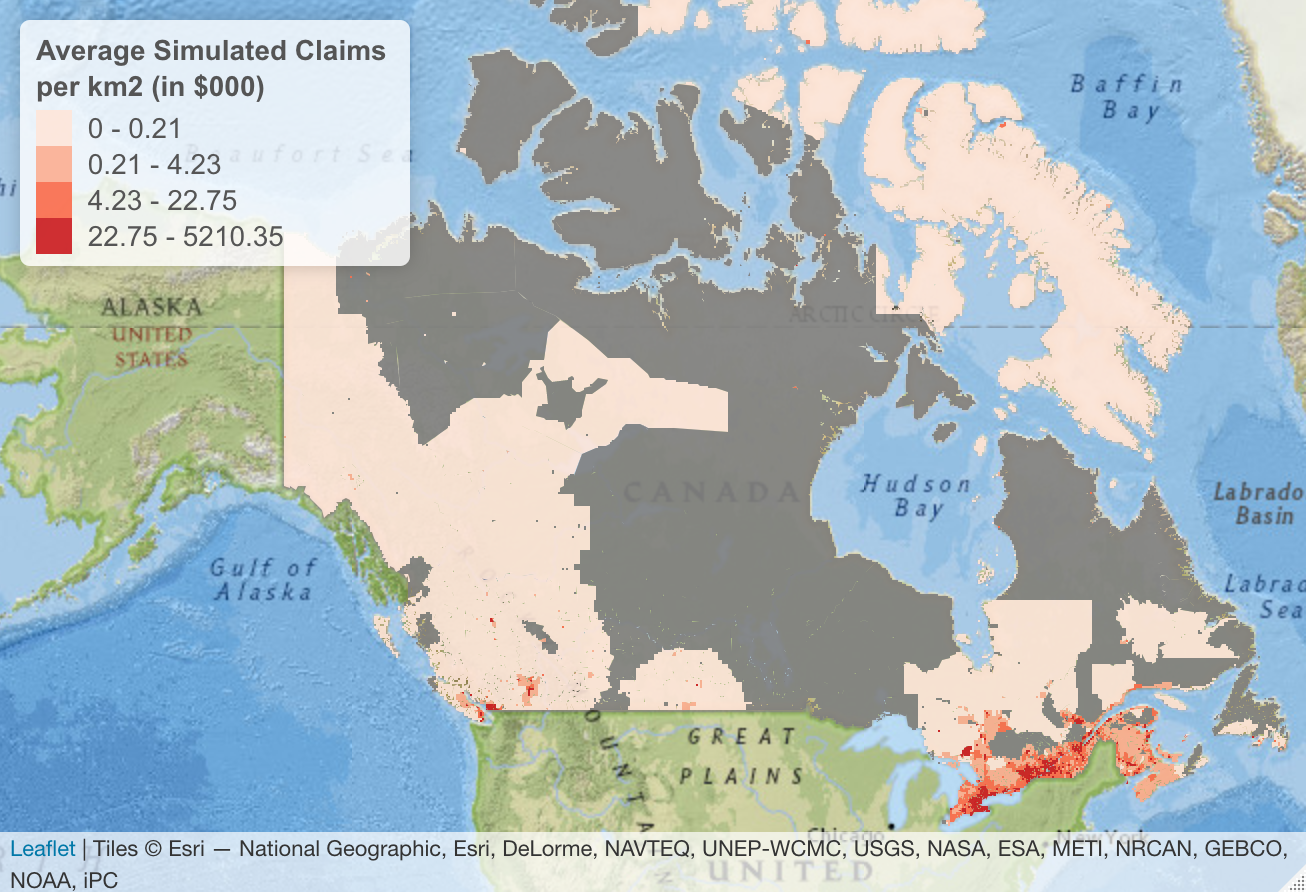}
  \caption{Average Simulated Claims}
\end{subfigure}
\caption{Average financial losses and insurance claims, conditional on the occurrence of an earthquake, based on simulated 100,000 years.}
\label{Figure: SimulatedLossesClaimsConditional}
\end{figure}

\begin{figure}[!ht]
    \centering
    \includegraphics[width=\textwidth]{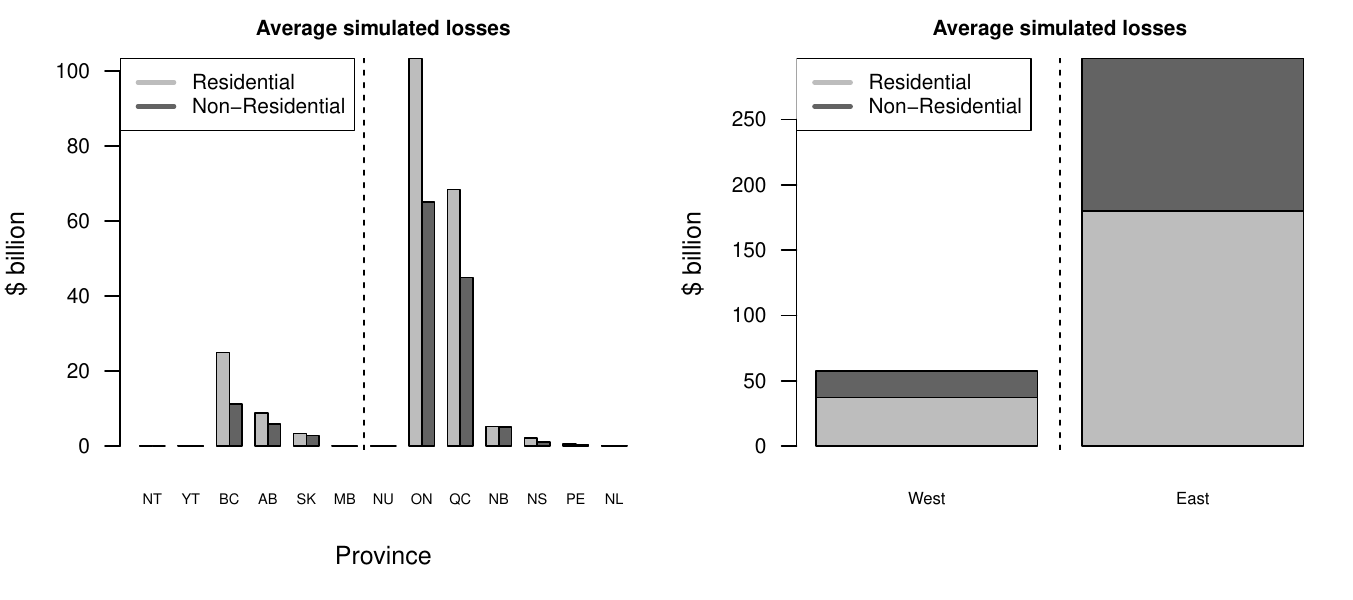}
    \caption{Average financial losses per province, conditional on the occurrence of an earthquake, based on 100,000 simulated years. The vertical dashed line splits Eastern and Western provinces.}
    \label{Figure: ExpectedLossesConditional}
\end{figure}

\begin{figure}[!ht]
    \centering
    \includegraphics[width=\textwidth]{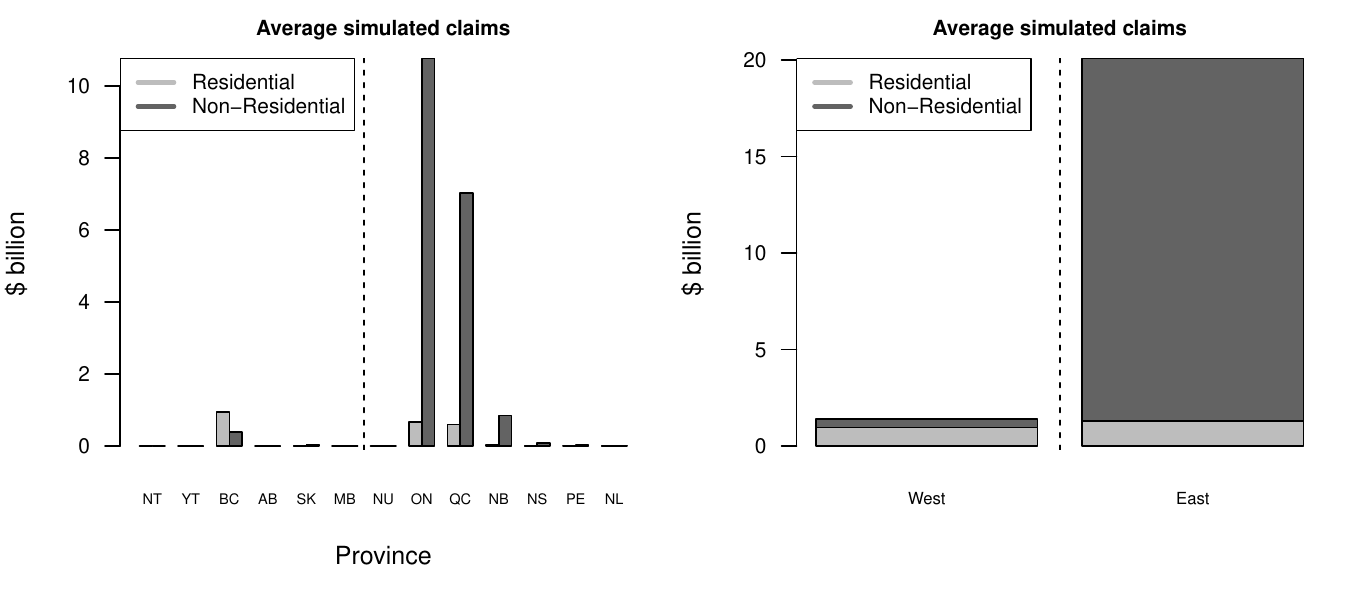}
    \caption{Average insurance claims per province, conditional on the occurrence of an earthquake,  based on 100,000 simulated years. The vertical dashed line splits Eastern and Western provinces.}
    \label{Figure: ExpectedClaimsConditional}
\end{figure}

Following the methodology explained in Section \ref{Section: Exposure} to calculate exposure and Algorithms \ref{Algorithm: EQSimulation} and \ref{Algorithm: EQLossesClaims}, we obtain values for financial losses and insurance claims for each simulated earthquake. Figure \ref{Figure: SimulatedLossesClaimsConditional} illustrates the expected value of the size of the simulated financial losses and insurance claims for each CSD, conditioning on the occurrence of an earthquake, whereas Figures \ref{Figure: ExpectedLossesConditional} and \ref{Figure: ExpectedClaimsConditional} compare the expected value per province. Some CSDs are located in zones free of any seismic activities such that they do not witness any significant earthquakes in the 100,000 simulated years and accordingly they observe no damage. The expected value of financial losses, conditional on the occurrence of an earthquake, for each province is relatively proportionate and consistent with the exposure values provided in Figure \ref{Figure: ExposureComparison}. Locations with high value of exposure witness high values of financial losses, however, insurance claims are affected by the insurance market penetration and the insurance terms. Given the large proportion of earthquake insurance market penetration for non-residential buildings in Eastern Canada compared to residential buildings, we observe a surge in the expected value of insurance claims for non-residential buildings. Larger losses are seen in Eastern Canada due to the predicted distance attenuation formulas, presented in eq.\ \eqref{Eq: EastBakun} and eq.\ \eqref{Eq: WestBakun}. For an earthquake of moment magnitude 6, MMI \rom{6} can be reached at a distance of 200 km from the epicenter in Eastern Canada, compared to a distance of 33 km from the epicenter in Western Canada. As discussed in \citet{EastWestUSGS}, seismic energy travels in Eastern North America much further than in Western North America.


In addition to the PML for the insurance claims, we estimate the Canada-wide PML for financial losses, which include the insured and non-insured portions of the losses, to provide a big picture of the damage. The correlation of the financial losses and insurance claims between the provinces is required for eq.\ \eqref{Eq: CorrPML}. Table \ref{Table: LossesCorrelation} and Tables \ref{Table: ClaimsCorrelation}, \ref{Table: LossesCorrelationKendall} and \ref{Table: ClaimsCorrelationKendall} in Appendix \ref{Appendix: CorrelationLossesClaims} provide Pearson's correlation coefficients of the financial losses, Pearson's correlation coefficients of the insurance claims, Kendall's tau of the financial losses and Kendall's tau of the insurance claims, respectively.
  
\begin{table}[!ht]
\centering\captionof{table}{Pearson correlation of the simulated financial losses between Canadian provinces, based on 100,000 years of simulated earthquakes.}
\label{Table: LossesCorrelation}
\begin{tabular}{|r|rrrrrrrrrrrrr|}
  \hline
 & \textbf{NL} & \textbf{PE} & \textbf{NS} & \textbf{NB} & \textbf{QC} & \textbf{ON} & \textbf{MB} & \textbf{SK} & \textbf{BC} & \textbf{YT} & \textbf{NT} & \textbf{AB} & \textbf{NU} \\ 
  \hline
\textbf{NL} & 1.00 & 0.38 & 0.32 & 0.31 & \multicolumn{1}{|c}{0.00} & 0.00 & 0.00 & 0.00 & 0.00 & 0.00 & 0.00 & 0.00 & 0.00 \\ 
  \textbf{PE} & 0.38 & 1.00 & 0.81 & 0.91 & \multicolumn{1}{|c}{0.00} & 0.00 & 0.00 & 0.00 & 0.00 & 0.00 & 0.00 & 0.00 & 0.00 \\ 
  \textbf{NS} & 0.32 & 0.81 & 1.00 & 0.82 & \multicolumn{1}{|c}{0.00} & 0.00 & 0.00 & 0.00 & 0.00 & 0.00 & 0.00 & 0.00 & 0.00 \\ 
  \cline{6-6}
  \textbf{NB} & 0.31 & 0.91 & 0.82 & 1.00 & 0.03 & \multicolumn{1}{|c}{0.00} & 0.00 & 0.00 & 0.00 & 0.00 & 0.00 & 0.00 & 0.00 \\ 
  \cline{2-4} \cline{7-8}
  \textbf{QC} & 0.00 & 0.00 & \multicolumn{1}{c|}{0.00} & 0.03 & 1.00 & 0.69 & 0.64 & \multicolumn{1}{|c}{0.00} & 0.00 & 0.00 & 0.00 & 0.00 & 0.00 \\ \cline{5-5} 
  \textbf{ON} & 0.00 & 0.00 & 0.00 & \multicolumn{1}{c|}{0.00} & 0.69 & 1.00 & 0.64 & \multicolumn{1}{|c}{0.00} & 0.00 & 0.00 & 0.00 & 0.00 & 0.00 \\ \cline{9-14}
  \textbf{MB} & 0.00 & 0.00 & 0.00 & \multicolumn{1}{c|}{0.00} & 0.64 & 0.64 & 1.00 & 0.03 & 0.00 & 0.03 & 0.02 & 0.00 & 0.02 \\ \cline{6-7}
  \textbf{SK} & 0.00 & 0.00 & 0.00 & 0.00 & 0.00 & \multicolumn{1}{c|}{0.00} & 0.03 & 1.00 & 0.04 & 0.02 & 0.02 & 0.08 & 0.02 \\ 
  \textbf{BC} & 0.00 & 0.00 & 0.00 & 0.00 & 0.00 & \multicolumn{1}{c|}{0.00} & 0.00 & 0.04 & 1.00 & 0.66 & 0.53 & 0.80 & 0.55 \\ 
  \textbf{YT} & 0.00 & 0.00 & 0.00 & 0.00 & 0.00 & \multicolumn{1}{c|}{0.00} & 0.03 & 0.02 & 0.66 & 1.00 & 0.87 & 0.40 & 0.88 \\ 
  \textbf{NT} & 0.00 & 0.00 & 0.00 & 0.00 & 0.00 & \multicolumn{1}{c|}{0.00} & 0.02 & 0.02 & 0.53 & 0.87 & 1.00 & 0.32 & 0.97 \\ 
  \textbf{AB} & 0.00 & 0.00 & 0.00 & 0.00 & 0.00 & \multicolumn{1}{c|}{0.00} & 0.00 & 0.08 & 0.80 & 0.40 & 0.32 & 1.00 & 0.35 \\ 
  \textbf{NU} & 0.00 & 0.00 & 0.00 & 0.00 & 0.00 & \multicolumn{1}{c|}{0.00} & 0.02 & 0.02 & 0.55 & 0.88 & 0.97 & 0.35 & 1.00 \\ 
  \hline
\end{tabular}
\end{table}  

Table \ref{Table: ParametersEstimates} shows the parameter estimates of the fitted marked homogeneous Poisson process with rate $\lambda$, shape parameter $\xi$ and scale parameter $\sigma$ for the size of the excesses over a high threshold $u$. The parameters and their standard errors (s.e.) are estimated from the simulated financial losses and insurance claims for each Canadian province, for Eastern and Western Canada and for the whole country. The thresholds are chosen to be the 0.95 quantile of the simulated data, or the 0.9 quantile for provinces with a small number of exceedances. Diagnostic plots (not shown) are used to confirm the adequacy of the fitted GPD models.

Tables \ref{Table: PMLLosses} and \ref{Table: PMLClaims} summarize the PML$_{1/x}$ for $x\in\lbrace100,250,500,750,1000\rbrace$ of the direct financial losses and the insurance claims, respectively. Results are provided for each province and for Eastern and Western Canada. The PML$_{1/x}$ are computed following two methods: by calculating the appropriate quantiles from the simulated data for each province (simulated), and by plugging in the parameter estimates from Table \ref{Table: ParametersEstimates} in eq.\ \eqref{Eq: PMLQuantile} (estimated). Additionally, Canada-wide PML$_{1/x}$ is calculated in three ways: following OSFI's formula in eq.\ \eqref{Eq: OSFIPML}, and the proposed formula in eq.\ \eqref{Eq: CorrPML} calculated with Pearson's correlation coefficient and Kendall's tau. We observe that the simulated and the estimated results are very similar, confirming the adequacy of the fitted models. We also observe that the proposed MCT formula in eq.\ \eqref{Eq: CorrPML} produces values that are comparable to OSFI's current approach in eq.\ \eqref{Eq: OSFIPML} and that results using Kendall's tau are more conservative than results using Pearson correlation. This is be explained by the larger values for Kendall's tau compared to Pearson Correlation in Eastern Canadian provinces, more specifically for QC.

\begin{figure}[!ht]
    \centering
    \includegraphics[width=\textwidth]{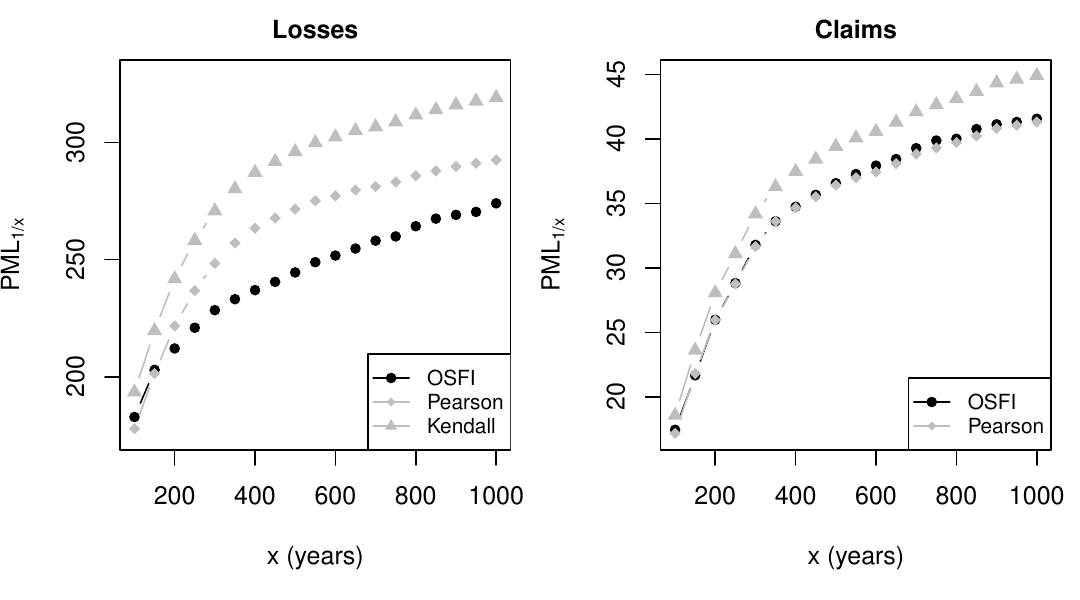}
    \caption{Canada-wide PML$_{1/x}$: OSFI vs eq.\ \eqref{Eq: CorrPML} with Pearson's correlation coefficient and Kendall's tau}
    \label{Figure: OSFI1inxYears}
\end{figure}

Figure \ref{Figure: OSFI1inxYears} compares the Canada-wide PML$_{1/x}$ for $x\in [100,1000]$ by using eq.\ \eqref{Eq: OSFIPML} and eq.\ \eqref{Eq: CorrPML} with Pearson correlation and with Kendall's tau. We observe that for the financial losses, the proposed formula is more conservative than OSFI's for any value of $x$, where the difference gap increases as $x$ increases, with Kendall's tau producing higher values than Pearson's correlation coefficient. For PML$_{1/500}$, eq.\ \eqref{Eq: CorrPML} with Pearson correlation is \$27 billion more than OSFI's formula, compared to a \$51 billion difference when Kendall's tau is used. This is explained by the fact that the proposed method captures the strong dependence between QC and ON, where both provinces are affected concurrently by the same earthquakes due to the distance attenuation equations in Eastern Canada producing larger isoseismal maps than Western Canada. However, for the PML$_{1/x}$ of insurance claims, eq.\ \eqref{Eq: CorrPML} with Pearson correlation coefficient produces very similar results to OSFI's, while Kendall's tau offers slightly higher values. In fact, for PML$_{1/500}$, eq.\ \eqref{Eq: CorrPML} with Kendall's tau is higher by approximately \$3 billion. Should Canadian P\&C insurance companies implement the proposed equation, especially with Pearson correlation coefficient, the transition will be smooth due to the resemblance in the values between both methods.

\begin{longsidewaystable}{|l|rrr|rrr|}
  \caption{Estimates of the parameters of the fitted homogeneous Poisson process for the simulated financial losses and insurance claims, based on 100,000 simulated years.} \label{Table: ParametersEstimates}
  \\ \hline
  \multirow{ 2}{*}{\textbf{Province}} & \multicolumn{3}{c|}{\textbf{Financial losses}} & \multicolumn{3}{c|}{\textbf{Insurance claims}}\\
  \cline{2-7}
 & $\sigma$ (s.e.) & $\xi$ (s.e.) & $\lambda$ (s.e.) &$\sigma$ (s.e.) &  $\xi$ (s.e.) & $\lambda$ (s.e.) \\ 
  \hline
  \textbf{NL} & 0.0836 (0.0067) & 0.0032 (0.0618) & 0.0040 (0.0002) & 0.0052 (0.0005) & 0.3388 (0.0890) & 0.0030 (0.0002) \\
  \textbf{PE} & 0.1832 (0.0197) & -0.0637 (0.0794) & 0.0019 (0.0001) & 0.0254 (0.0025) & 0.2751 (0.0836) & 0.0036 (0.0002) \\
  \textbf{NS} & 1.5970 (0.1376) & -0.2589 (0.0556) & 0.0022 (0.0001) & 0.1369 (0.0133) & 0.0109 (0.0667) & 0.0020 (0.0001) \\  
  \textbf{NB} & 2.3892 (0.1197) & -0.1398 (0.0337) & 0.0072 (0.0003) & 0.7719 (0.0424) & -0.0566 (0.0399) & 0.0070 (0.0003) \\ 
  \textbf{QC} & 31.5666 (1.2992) & -0.1804 (0.0262) & 0.0096 (0.0003) & 11.1067 (0.4390) & -0.2301 (0.0238) & 0.0095 (0.0003) \\ 
  \textbf{ON} & 38.4570 (1.4825) & -0.1784 (0.0197) & 0.0088 (0.0003) & 2.4030 (0.1313) & 0.2751 (0.0437) & 0.0086 (0.0003) \\
  \textbf{MB} & 0.4009 (0.0166) & -0.3343 (0.0200) & 0.0067 (0.0003) & 0.1173 (0.0051) & -0.3647 (0.0208) & 0.0051 (0.0002) \\ 
  \textbf{SK} & 0.0801 (0.0107) & 1.6201 (0.1526) & 0.0030 (0.0002) & 0.0042 (0.0005) & 0.8859 (0.1369) & 0.0017 (0.0001) \\ 
  \textbf{BC} & 2.8283 (0.0800) & -0.1021 (0.0174) & 0.0200 (0.0004) & 0.1170 (0.0056) & 0.6930 (0.0444) & 0.0157 (0.0004) \\
  \textbf{YT} & 0.9433 (0.0620) & -0.1498 (0.0434) & 0.0040 (0.0002) & 0.0903 (0.0118) & 0.6248 (0.1217) & 0.0026 (0.0002) \\ 
  \textbf{NT} & 6.8276 (0.5989) & -0.0433 (0.0609) & 0.0025 (0.0002) & 1.4643 (0.1392) & 0.2565 (0.0770) & 0.0030 (0.0002) \\ 
  \textbf{AB} & 0.2505 (0.0207) & 0.2880 (0.0638) & 0.0033 (0.0002) & 0.1236 (0.0128) & -0.0426 (0.0764) & 0.0020 (0.0001) \\
  \textbf{NU} & 2.0897 (0.1620) & -0.1284 (0.0458) & 0.0025 (0.0002) & 0.0988 (0.0088) & 0.7871 (0.0841) & 0.0049 (0.0002) \\  
  \textbf{East} & 38.9950 (1.4906) & -0.0827 (0.0222) & 0.0103 (0.0003) & 11.9035 (0.4675) & -0.1046 (0.0238) & 0.0101 (0.0003) \\  
  \textbf{West} & 11.8276 (0.4373) & 0.0851 (0.0309) & 0.0243 (0.0005) & 0.1912 (0.0088) & 0.9903 (0.0462) & 0.0198 (0.0004) \\  
  \textbf{Total} & 39.5434 (1.3060) & -0.0752 (0.0206) & 0.0149 (0.0004) & 12.6883 (0.5151) & -0.1385 (0.0194) & 0.0077 (0.0003) \\ 
   \hline
\end{longsidewaystable}   

\begin{longsidewaystable}{|r|r|p{0.6cm}p{0.6cm}p{0.6cm}p{0.6cm}p{0.9cm}p{0.9cm}p{0.6cm}p{0.6cm}p{0.6cm}p{0.6cm}p{0.6cm}p{0.6cm}p{0.6cm}p{1cm}p{0.8cm}|p{1cm}p{1.1cm}p{1.3cm}|}
\caption{PML$_{1/x}$ values (in \$ billions) for the financial losses, based on 100,000 simulated years. }
\label{Table: PMLLosses}
 \\ \hline
&x & NL & PE & NS & NB & QC & ON & MB & SK & BC & YT & NT & AB & NU & East & West & OSFI & Pearson & Kendall \\ 
  \hline
\parbox[t]{2mm}{\multirow{5}{*}{\rotatebox[origin=c]{90}{\textbf{Simulated}}}}&100 & 0.1 & 0.9 & 2.6 & 10.4 & 135.0 & 53.4 & 0.8 & 0.0 & 4.0 & 0.9 & 9.7 & 0.4 & 1.3 & 180.1 & 14.9 & 182.9 & 177.9 & 193.6 \\ 
  &250 & 0.1 & 1.4 & 4.3 & 12.6 & 164.6 & 87.3 & 1.1 & 0.1 & 6.2 & 1.7 & 19.5 & 0.7 & 3.8 & 214.2 & 28.1 & 221.0 & 236.8 & 258.2 \\ 
  &500 & 0.2 & 1.6 & 5.6 & 13.6 & 180.5 & 108.6 & 1.2 & 0.1 & 7.7 & 2.4 & 25.7 & 0.9 & 5.4 & 234.4 & 38.1 & 244.6 & 271.6 & 296.0 \\ 
  &750 & 0.2 & 1.7 & 6.2 & 14.2 & 185.7 & 115.4 & 1.3 & 0.2 & 8.5 & 2.7 & 28.5 & 1.0 & 6.3 & 248.4 & 42.3 & 259.9 & 283.1 & 308.8 \\ 
  &1000 & 0.3 & 1.7 & 6.6 & 14.7 & 189.7 & 121.3 & 1.3 & 0.3 & 9.1 & 2.9 & 30.6 & 1.1 & 6.6 & 261.6 & 45.4 & 274.0 & 292.5 & 319.1 \\ 
 \hline
\parbox[t]{2mm}{\multirow{5}{*}{\rotatebox[origin=c]{90}{\textbf{Estimated}}}}&100 & 0.1 & 1.3 & 2.5 & 10.3 & 134.4 & 52.7 & 0.7 & 0.0 & 3.9 & 0.8 & 14.4 & 0.5 & 1.6 & 179.9 & 13.9 & 182.4 & 177.2 & 192.7 \\ 
  &250 & 0.1 & 1.5 & 4.4 & 12.5 & 161.4 & 86.1 & 1.0 & 0.1 & 6.2 & 1.7 & 20.9 & 0.7 & 3.8 & 214.3 & 26.1 & 220.4 & 232.9 & 254.0 \\ 
  &500 & 0.2 & 1.6 & 5.6 & 13.9 & 179.0 & 107.9 & 1.2 & 0.1 & 7.9 & 2.4 & 25.7 & 0.9 & 5.3 & 238.7 & 36.0 & 247.9 & 269.6 & 294.2 \\ 
  &750 & 0.2 & 1.7 & 6.2 & 14.7 & 188.3 & 119.5 & 1.3 & 0.2 & 8.7 & 2.7 & 28.4 & 1.0 & 6.1 & 252.3 & 42.0 & 263.6 & 289.2 & 315.6 \\ 
  &1000 & 0.2 & 1.7 & 6.6 & 15.3 & 194.5 & 127.2 & 1.4 & 0.3 & 9.4 & 2.9 & 30.3 & 1.1 & 6.6 & 261.6 & 46.4 & 274.5 & 302.2 & 329.8 \\ 
   \hline
   \multicolumn{1}{c}{}&\multicolumn{19}{l}{Simulated: calculated by solving for the appropriate quantiles in simulated data for each province.}\\
   \multicolumn{1}{c}{}&\multicolumn{19}{l}{Estimated: calculated by plugging in the parameters estimates from Table \ref{Table: ParametersEstimates} in eq.\ \eqref{Eq: PMLQuantile}.}\\
\end{longsidewaystable}

\begin{longsidewaystable}{|r|r|p{0.6cm}p{0.6cm}p{0.6cm}p{0.6cm}p{0.9cm}p{0.9cm}p{0.6cm}p{0.6cm}p{0.6cm}p{0.6cm}p{0.6cm}p{0.6cm}p{0.6cm}p{1cm}p{0.8cm}|p{1cm}p{1.1cm}p{1.3cm}|}
\caption{PML$_{1/x}$ values (in \$ billions) for the insurance claims, based on 100,000 simulated years.}\label{Table: PMLClaims}
\\  \hline
&x & NL & PE & NS & NB & QC & ON & MB & SK & BC & YT & NT & AB & NU & East & West & OSFI & Pearson & Kendall \\ 
  \hline
\parbox[t]{2mm}{\multirow{5}{*}{\rotatebox[origin=c]{90}{\textbf{Simulated}}}}&100 & 0.0 & 0.0 & 0.1 & 0.9 & 12.7 & 5.6 & 0.1 & 0.0 & 0.1 & 0.0 & 0.0 & 0.0 & 0.0 & 17.4 & 0.3 & 17.4 & 17.2 & 18.6 \\ 
  &250 & 0.0 & 0.1 & 0.2 & 1.6 & 22.3 & 8.2 & 0.2 & 0.0 & 0.3 & 0.0 & 0.3 & 0.0 & 0.1 & 28.7 & 1.1 & 28.8 & 28.8 & 31.1 \\ 
  &500 & 0.0 & 0.1 & 0.3 & 2.1 & 28.2 & 10.3 & 0.2 & 0.0 & 0.7 & 0.1 & 1.2 & 0.1 & 0.2 & 36.3 & 2.0 & 36.6 & 36.4 & 39.4 \\ 
  &750 & 0.0 & 0.1 & 0.4 & 2.3 & 30.0 & 11.7 & 0.2 & 0.0 & 0.9 & 0.1 & 1.8 & 0.1 & 0.3 & 39.4 & 2.9 & 39.9 & 39.3 & 42.6 \\ 
  &1000 & 0.0 & 0.1 & 0.4 & 2.5 & 31.1 & 12.7 & 0.2 & 0.0 & 1.2 & 0.2 & 2.6 & 0.1 & 0.4 & 40.9 & 3.4 & 41.6 & 41.3 & 44.9 \\ 
   \hline
 \parbox[t]{2mm}{\multirow{5}{*}{\rotatebox[origin=c]{90}{\textbf{Estimated}}}}&  100 & 0.0 & 0.1 & 0.1 & 0.9 & 12.4 & 5.8 & 0.0 & 0.0 & 0.1 & 0.0 & 0.0 & 0.0 & 0.0 & 17.5 & 0.3 & 17.5 & 17.0 & 18.4 \\ 
  &250 & 0.0 & 0.1 & 0.2 & 1.6 & 21.7 & 8.2 & 0.1 & 0.0 & 0.3 & 0.0 & 0.2 & 0.0 & 0.1 & 27.9 & 0.9 & 28.0 & 28.2 & 30.5 \\ 
  &500 & 0.0 & 0.1 & 0.3 & 2.1 & 27.6 & 10.5 & 0.2 & 0.0 & 0.6 & 0.1 & 1.2 & 0.1 & 0.2 & 35.1 & 1.8 & 35.4 & 35.9 & 38.9 \\ 
  &750 & 0.0 & 0.1 & 0.4 & 2.4 & 30.6 & 12.0 & 0.2 & 0.0 & 0.8 & 0.1 & 1.9 & 0.1 & 0.3 & 39.1 & 2.7 & 39.6 & 40.2 & 43.6 \\ 
  &1000 & 0.0 & 0.1 & 0.4 & 2.6 & 32.5 & 13.2 & 0.3 & 0.0 & 1.0 & 0.2 & 2.5 & 0.1 & 0.4 & 41.9 & 3.6 & 42.6 & 43.2 & 46.9 \\ 
   \hline
   \multicolumn{1}{c}{}&\multicolumn{19}{l}{Simulated: calculated by solving for the appropriate quantiles in simulated data for each province.}\\
   \multicolumn{1}{c}{}&\multicolumn{19}{l}{Estimated: calculated by plugging in the parameters estimates from Table \ref{Table: ParametersEstimates} in eq.\ \eqref{Eq: PMLQuantile}.}\\
\end{longsidewaystable} 

\section{Discussion}\label{Section: Discussion}
Canada has elevated seismic risk due to the presence of urban population in zones with high seismic activities, such as in Eastern and Western Canada. Estimation of the financial damage is relevant to insurance companies, governmental institutions and homeowners. Assessment of the spatio-temporal models by means of goodness-of-fit tests can be performed through residual analysis methods. In this paper, we fit two different models to the sptio-temporal point pattern for significant Canadian earthquakes. We extend the use of residual analysis on Voronoi polygons by introducing deviance Voronoi residuals, which provide a very useful tool to compare fitted models and identify locations where one model is superior to the other. We also create an earthquake financial losses estimation tool for Canada by relying on building information, their replacement costs and earthquake damage probability matrices. Additionally, insurance policy terms and market information are used to estimate insurance claim values. A more interpretable approach is suggested to calculate the county-wide PML by relying on the correlation between neighboring provinces. A large simulation of 100,000 years of earthquakes is performed, where we obtain parameter estimates for the tail behavior by using extreme value theory techniques. The results displayed in this article can be improved further by obtaining data on the non-residential buildings in Canada and more detailed information on earthquake insurance penetration and policy terms in Canadian municipalities. Based on our methodology and parameters, a significant earthquake that occurs at a rate of 1-in-500 years in Qu\'ebec can cause financial damages of around \$180 billion. Yet, earthquake insurance claims that occur at a rate of 1-in-500 years are only around \$28 billion. Some of the \$152 billion in uncovered losses can be partially covered by homeowners, but the government may have to intervene, especially to repair damages in infrastructure. This represents considerable government expenditures, for example, it is nine-fold the budgeted COVID-19 support and recovery measures for the years 2020-2024. Covering \$152 billion would double the planned total expenditures for 2021-2022 \autocite{BudgetQC2021} and lead to a major deficit. Analogous comparisons show that governments in Ontario and other provinces are also ill-prepared for earthquake relief and that there is a need for further insurance market penetration, especially in Eastern Canada.

\section*{Supplementary Material}\label{Section: ShinyApp}

An open-source interactive web application is provided in \url{https://anonymoususer.shinyapps.io/EQSimulator/}. The application allows the user to simulate multiple significant earthquakes, with different random moment magnitudes, in a chosen geographical location in Canada. The insurance market penetration and insurance terms are chosen by the user and their resulting insurance claims are calculated. This web application can provide insurers with a simulated value of the expected financial losses in case of an occurrence of a significant earthquake in areas where they have exposure or plan to sell new earthquake policies. It also provides the simulated isoseismal map, which contains information on the exposure and losses for the affected CSDs.

\newpage
\printbibliography
\newpage

\begin{appendices}
\section{Modified Mercalli Intensity Definitions}
\label{Appendix: Tables MMI}

\begin{longtable}{|p{1.85cm}|p{1.65cm}|p{11.3cm}| }
\caption{Modified Mercalli Intensity Definitions \autocite{MMI}}\label{Table:MMI}\\
\hline
\textbf{Intensity} & \textbf{Shaking} & \textbf{Description}\\
\hline
\rom{1}& Not felt& Not felt except by very few under especially favorable conditions.\\
\hline
\rom{2}& Weak & Felt only by a few people at rest, especially on upper floors of buildings. Delicately suspended objects may swing.\\
\hline
\rom{3}& Weak & Felt quite noticeably by people indoors, especially on upper floors of buildings: Many people do not recognize it as an earthquake. Standing motor cars may rock slightly. Vibrations are similar to the passing of a truck.\\
\hline
\rom{4}& Light & Felt indoors by many, outdoors by few during the day: At night, some are awakened. Dishes, windows, and doors are disturbed; walls make cracking sounds. Sensations are like a heavy truck striking a building. Standing motor cars are rocked noticeably.\\
\hline
\rom{5}& Moderate & Felt by nearly everyone; many awakened: Some dishes and windows are broken. Unstable objects are overturned.\\
\hline
\rom{6}& Strong & Felt by all, and many are frightened. Some heavy furniture is moved; a few instances of fallen plaster occur. Damage is slight.\\
\hline
\rom{7}& Very strong & Damage is negligible in buildings of good design and construction; but slight to moderate in well-built ordinary structures; damage is considerable in poorly built or badly designed structures; some chimneys are broken. Noticed by people in driving motor cars.\\
\hline
\rom{8}& Severe & Damage slight in specially designed structures; considerable damage in ordinary substantial buildings with partial collapse. Damage great in poorly built structures. Fall of chimneys, factory stacks, columns, monuments, walls. Heavy furniture overturned. Sand and mud ejected in small amounts. Changes in well water. People in driving motor cars are disturbed.\\
\hline
\rom{9}& Violent&Damage is considerable in specially designed structures; well-designed frame structures are thrown out of plumb. Damage is great in substantial buildings, with partial collapse. Buildings are shifted off foundations. Liquefaction occurs. Underground pipes are broken.\\
\hline
\rom{10}& Extreme & Some well-built wooden structures are destroyed; most masonry and frame structures are destroyed with foundations. Rails are bent. Landslides considerable from river banks and steep slopes. Shifted sand and mud. Water splashed over banks.\\
\hline
\rom{11} & Extreme& Few, if any, (masonry) structures remain standing. Bridges are destroyed. Broad fissures erupt in the ground. Underground pipelines are rendered completely out of service. Earth slumps and land slips in soft ground. Rails are bent greatly.\\
\hline
\rom{12} & Extreme& Damage is total. Waves are seen on ground surfaces. Lines of sight and level are distorted. Objects are thrown upward into the air.\\
\hline
\end{longtable}

\newpage
\section{Collection of building inventory and calculation of exposure}  \label{Section: Appendix Exposure}

We define:
\begin{itemize}
             \item Total square footage = \# units in a building type $\times$ average square footage of that type of unit
             \item Building exposure = Total square footage $\times$ Mean replacement cost
             \item Building content exposure =  Building exposure $\times$ \% contents value.
\end{itemize}
Now we explain the sources of the building inventory data as well as the repair and replacement cost information.

\textbf{\textit{Residential Dwellings}}\\
Information on the number of buildings for each residential building classification is obtained from \citet{statistics_Canada_2016}. The building classifications (single-detached houses, row houses, etc.) are transformed into HAZUS' occupancy codes as per Table C.4 in \citet{Canada_HAZUS}. The Canadian Housing Statistics Program contains comprehensive data on the average total living area in square feet for each residential building class in each CSD \autocite{statistics_Canada_2020_housing}. The data are currently available for BC, ON, NS and NB. For CSDs that have missing square footage values, we use the square footage of the CSD in the same province that has the closest household median income value, obtained from \citet{statistics_Canada_2016}. For CSDs that have missing square footage values and missing household median income value, we use the weighted average square footage of the province, where the weights are the number of houses in each CSD. For the provinces that do not have square footage values, we use the values of a CSD that has the nearest household income values, where the provinces are matched as follows:
\begin{itemize}
    \item ON square footage is used for for QC, MB, and NU.
    \item NB square footage is used for NL and PE.
    \item BC square footage is used for SK, AB, YT, and NT.
\end{itemize}

For dwelling types that do not have square footage information, we assume that an \quotes{other single-attached} house has the same square footage as a \quotes{single-detached} house and an \quotes{apartment or flat in a duplex} has the same square footage as a \quotes{Semi-detached} house. We use the values suggested in \citet{Canada_HAZUS} for the remaining dwelling types that do not have information on square footage.

The building replacement cost, in dollar units, for each HAZUS occupancy code are obtained from HAZUS \citet{Canada_HAZUS}, which was originally obtained from the RSMeans. The building construction price index is used to inflate the construction costs to June 2021 \autocite{Statscan_BCPI}. The BCPI is available for eleven census metropolitan areas, thus we generalize the inflation rate to each area's respective province. Some provinces/territories were not included in the data and thus we assumed that NL, NU, NT, YT, PE, SK and MB will follow the smallest inflation value, which is that of Edmonton, AB. We also assumed that NB will follow the inflation trend of Halifax, NS.

The building content replacement value for residential dwellings is assumed to be at $50\%$ of the building replacement cost as suggested in \citet{Canada_HAZUS, federal2013multi}. The building exposure for each HAZUS code is converted to construction types (wood, concrete, steel, masonry, etc.) by using HAZUS' general building scheme mapping information, available in Table 5.1 in \citet{federal2013multi}.

\textbf{\textit{Non-residential Dwellings}}\\
Statistics Canada currently does not have a comprehensive dataset for non-residential buildings, thus, we will rely on their building permits data, available in \citet{Statscan_BuildingPermits}. The average annual ratios of institutional and governmental, commercial and industrial building permits to residential building permits are calculated over the years 2011 to 2019. Accordingly, the exposure of non-residential buildings is calculated as a percentage of the residential buildings total exposure. Additionally, a detailed split of each building category is available in \citet{Statscan_BuildingPermits}. Table \ref{Table:NonResiBuildings} provides the suggested conversion of non-residential building types to HAZUS occupancy codes.

Similar to residential buildings, the exposure by HAZUS dwelling type is converted to construction types (wood, concrete, steel, masonry, etc.), following Table 5.1 in \citet{federal2013multi}. The building content replacement value for non-residential dwellings is assumed to be a percentage of the building replacement cost, as suggested by \citet{federal2013multi}.

\begin{longtable}{|p{7.5cm}|p{2.3cm}|p{5cm}|}
\caption{Conversion from Statistics Canada classification to HAZUS occupancy codes for non-residential buildings}\label{Table:NonResiBuildings}\\
\hline
\textbf{Statistics Canada Building Permits Label} &\textbf{HAZUS Occupancy Code}&\textbf{Description}\\
\hline
\textbf{Institutional and governmental}&&\\
Elementary school, kindergarten	&EDU1&	Grade Schools\\
Secondary school, high school, junior high school&	EDU1&	Grade Schools\\
Post-secondary institution and technical institute&	EDU2&Colleges/Universities\\
University&	EDU2&	Colleges/Universities\\
Library, museum, art gallery, aquarium, botanical garden, scientific center&	COM8&	Entertainment, recreation\\
General hospital&	COM6&	Hospital\\
Clinic, out-patient clinic, first aid station	&COM7&	Medical Office/Clinic\\
Welfare, home&	RES5&	Institutional dormitory\\
Churches, religion&	REL1&	Church/Non-Profit\\
Government legislative and administration building, city hall, court of justice, embassy, parliament and senate building&	GOV1&	General services\\
Other government building - police station, prison, fire station, military building&	GOV2&	Emergency response\\
\hline
\textbf{Industrial		}&&\\
Maintenance building&	IND1&	Heavy factory\\
Plant for manufacturing, processing and assembling goods&	IND1&	Heavy factory\\
Communication building	&IND1&	Heavy factory\\
Transportation terminal&	IND1&	Heavy factory\\
Utility building&	IND1&	Heavy factory\\
Mining building	&IND4&	Metals/minerals processing\\
Agriculture	&AGR1&	Agriculture\\
\hline
\textbf{Commercial }&&\\
Trade and services&	COM1&	Retail trade\\
Warehouses&	COM2&	Wholesale trade\\
Service stations&	COM3&	Personal and repair services\\
Office buildings&	COM4&	Professional/technical services\\
Theatre and performing art center, movie theatre, concert hall, opera house, cultural center&	COM9&	Theaters\\
Indoor recreational building, sports complex, tennis court and squash, community center, arena, curling club, swimming pool&	COM8&	Entertainment, recreation\\
Outdoor recreational building, country club, golf club campground facilities, outdoor skating rink, outdoor swimming pool&	COM8&	Entertainment, recreation\\
Convention center, exhibition building&	COM4	&Professional/technical services\\
Hotel, hotel and motel, motor hotel&	RES4&	Temporary lodging\\
Motel, cabin for tourism&	RES4&	Temporary lodging\\
Student's residence, boarding house, religious residence, hostel, dormitory&	RES5&	Institutional dormitory\\
Restaurant, bar, night club, diner&	COM8&	Entertainment, recreation\\
Laboratories&	COM7&	Medical Office/Clinic\\
\hline
\end{longtable}

\section{Residual Analysis of the Fitted STPP Models}\label{Appendix: Section STPP}
\begin{figure}[H]
    \centering
    \includegraphics[width=\textwidth]{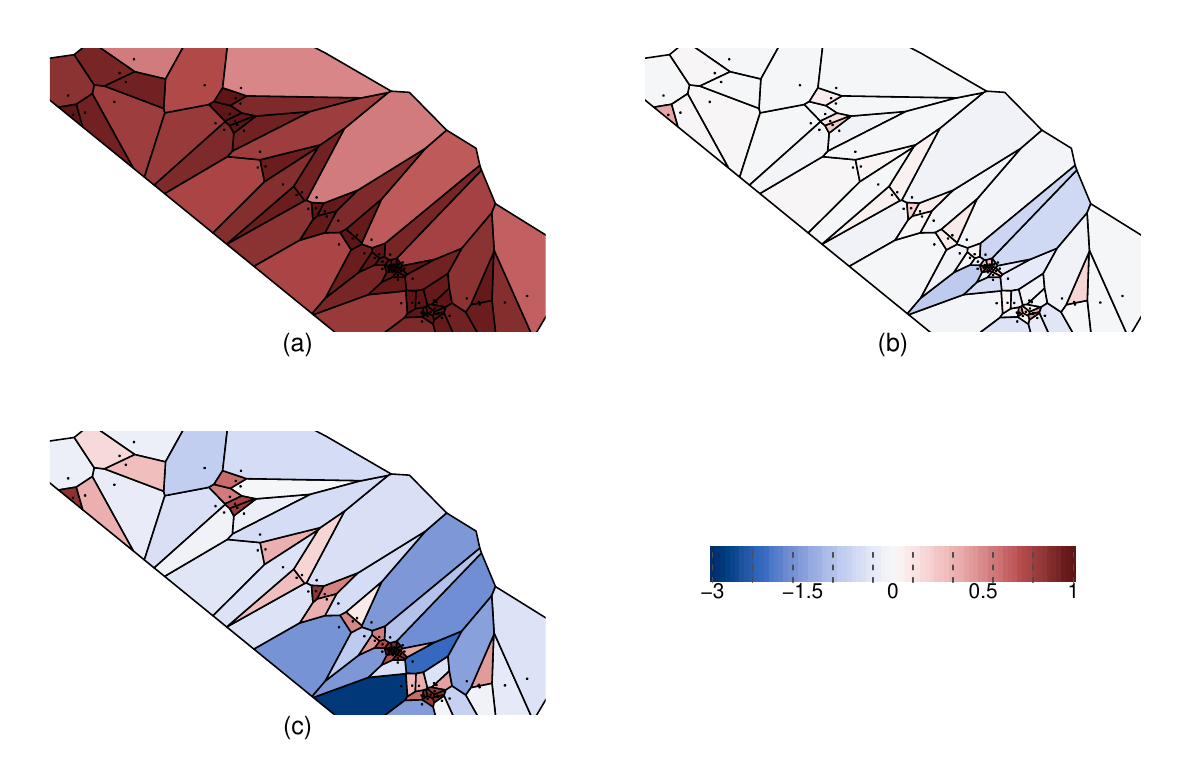}
    \caption{Focusing on Western Canada, raw Voronoi residuals of models $P$ (a), $H_1$ (b), and $H_2$ (c).}
    \label{Figure: RawWest}
\end{figure}
\begin{figure}[H]
    \centering
    \includegraphics[width=\textwidth]{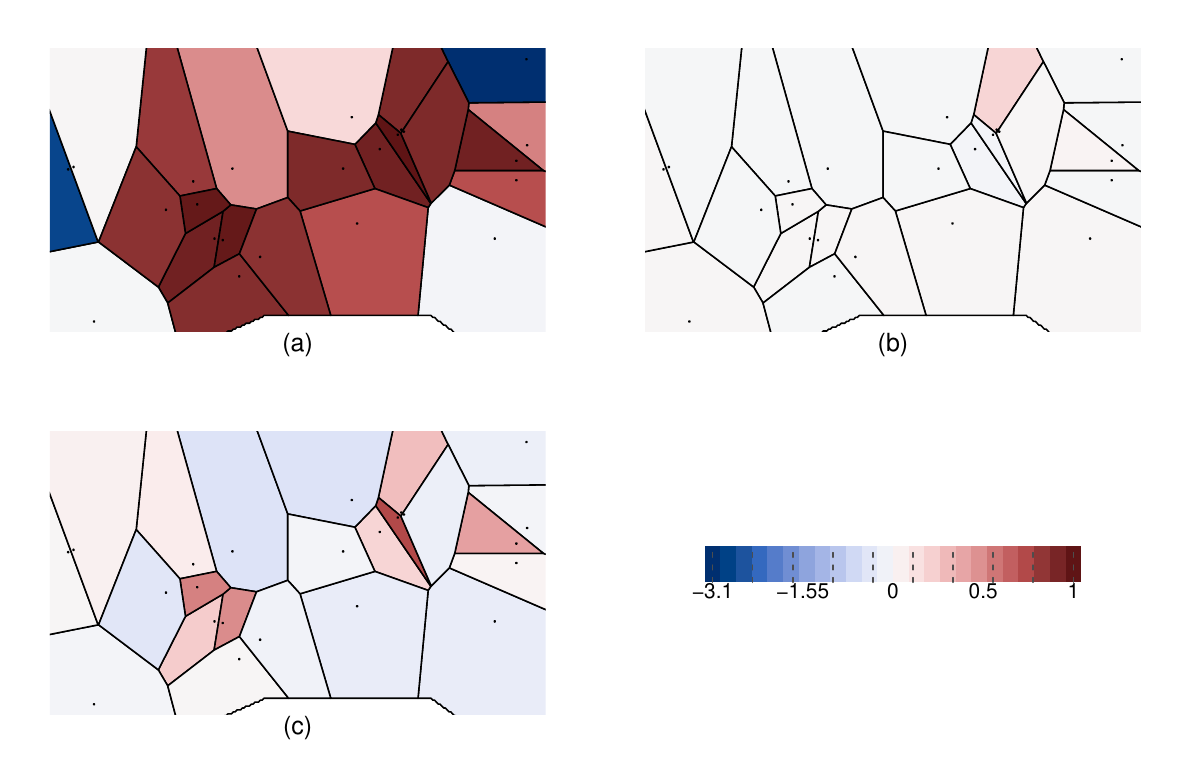}
    \caption{Focusing on Eastern Canada, raw Voronoi residuals of models $P$ (a), $H_1$ (b), and $H_2$ (c).}
    \label{Figure: RawEast}
\end{figure}
\begin{figure}[H]
    \centering
    \includegraphics[width=\textwidth]{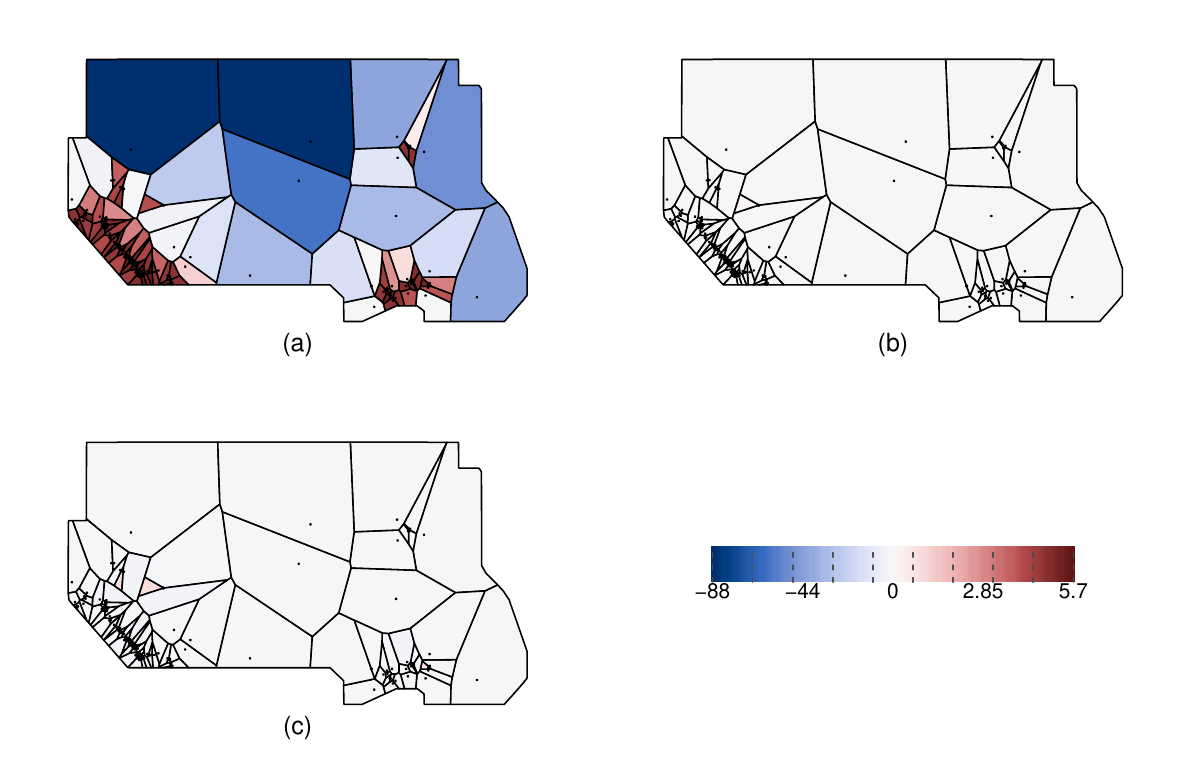}
    \caption{Pearson Voronoi residuals of models $P$ (a), $H_1$ (b), and $H_2$ (c).}
    \label{Figure: Pearson}
\end{figure}
\begin{figure}[H]
    \centering
    \includegraphics[width=\textwidth]{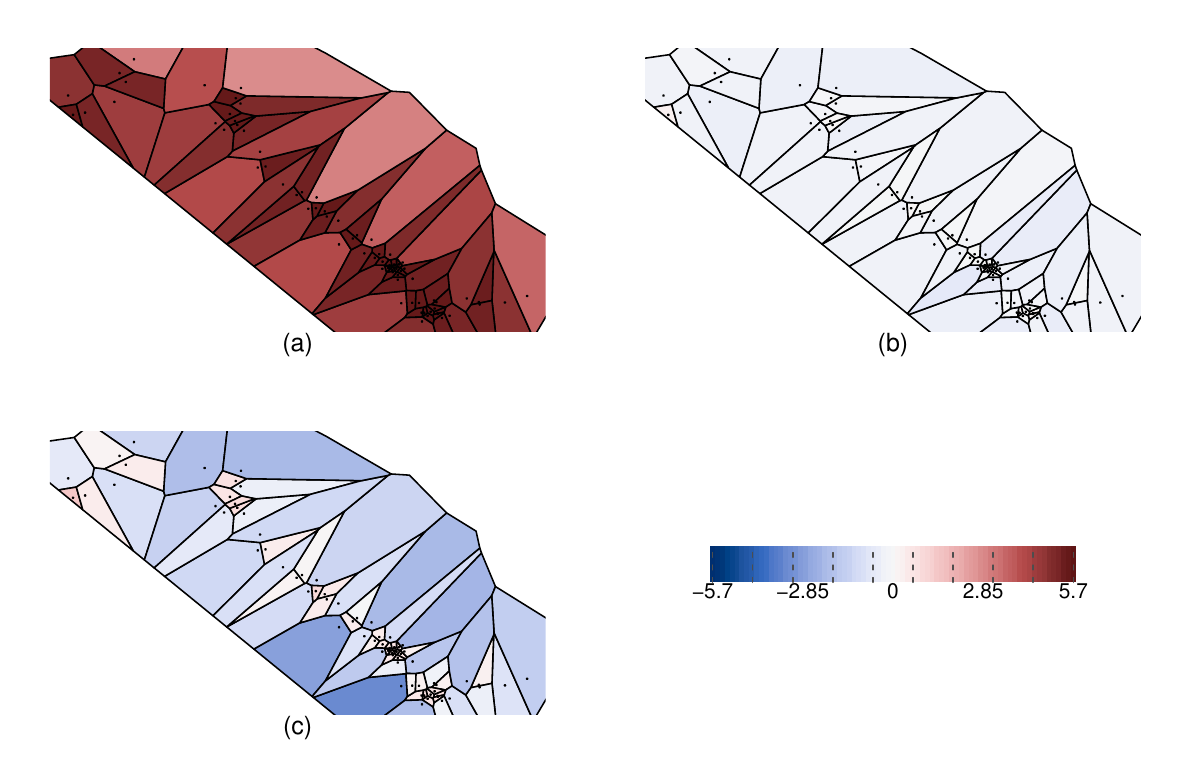}
    \caption{Focusing on Western Canada, Pearson Voronoi residuals of models $P$ (a), $H_1$ (b), and $H_2$ (c).}
    \label{Figure: PearsonWest}
\end{figure}
\begin{figure}[H]
    \centering
    \includegraphics[width=\textwidth]{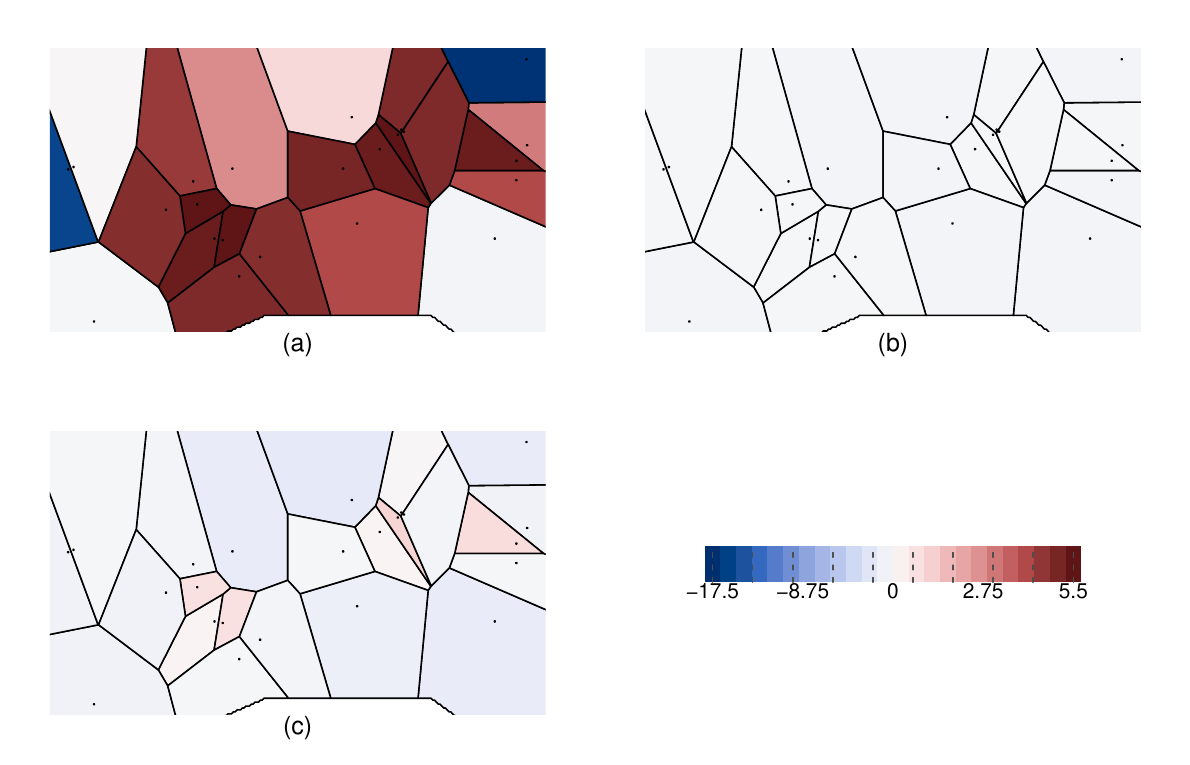}
    \caption{Focusing on Eastern Canada, Pearson Voronoi residuals of models $P$ (a), $H_1$ (b), and $H_2$ (c).}
    \label{Figure: PearsonEast}
\end{figure}

\newpage
\section{Extreme Value Theory} \label{Section: Appendix EVT}
Extreme value analysis is a branch in statistics that is focused on the behavior of the tail of the distribution \autocite{coles2001introduction}. There are two principle models for extreme values: the block maxima model and the peaks-over-threshold model. The \textit{block maxima} approach is used to model the largest observations from samples of identically distributed observations in successive blocks. The \textit{peaks-over-threshold} is used to model all large observations that exceed a given high threshold value, denoted $u$. 

\subsection{Distribution Of The Maxima}
The limiting distribution of block maxima is given in the following theorem:

\begin{Theorem}\textbf{Fisher and Tippet theorem} \autocite{fisher1928limiting} \label{Thm: FisherTippet}\\
Let $X_1, \ldots, X_n$ be a sequence of independent random variables having a common distribution function $F$ and consider $M_n = \max\lbrace X_1,\ldots,X_n\rbrace$. If there exists norming constants $(a_n)$ and $(b_n)$, where $a_n\in\mathbb{R}$ and $b_n>0$ for all $n\in\mathbb{N}$ and some non-degenerate distribution function $H$ such that $$\frac{M_n-a_n}{b_n}\xrightarrow{\text{d}} H,$$ then $H$ belongs to one of the following three classes of distributions (up to location and scaling):
\begin{align*}
\text{Fr\'echet:} \ \ \ &\Phi_\alpha(x) = \begin{cases} 
     0,   &\ \   x\leq0,\\
     \exp\left\lbrace-x^{-\alpha}\right\rbrace,    &\ \  x>0,\\
  \end{cases}\ \ \ \alpha>0\\
  \text{Gumbel:} \ \ \ &\Lambda(x) = \exp\left\lbrace-\exp\lbrace-x\rbrace\right\rbrace, \ \ \ x\in\mathbb{R},\\
  \text{Weibull:} \ \ \ &\Psi_\alpha(x) = \begin{cases} 
      \exp\left\lbrace-(-x)^{\alpha}\right\rbrace,   &\   x\leq0,\\
     1,    &\  x>0,\\
  \end{cases}\ \ \ \alpha>0
\end{align*}

\end{Theorem}

Results by \citet{VonMises1954, jenkinson1955frequency} provide a generalization of Theorem \ref{Thm: FisherTippet}. Set $\xi=\alpha^{-1}$ for the Fr\'echet distribution, $\xi=-\alpha^{-1}$ for the Weibull distribution and interpret the Gumbel distribution as a limiting case as $\xi\to0$, then we obtain the following definition.

\begin{Definition}{\textbf{Generalized Extreme Value Distribution}.}\\

The distribution function of a GEV is given by
\begin{align*}
H_{\xi}(x)&=
\begin{cases} 
      \exp \left\lbrace -\left(1+\xi x \right)^{-\frac{1}{\xi}}\right\rbrace,   &\ \   \xi\neq0,\\
      \exp\left\lbrace-\exp(-x)\right\rbrace,    &\ \  \xi=0,\\
  \end{cases}
\end{align*}
where $1+\xi x>0$. A three-parameter family is obtained by defining $H_{\xi,\mu,\sigma} :=H_{\xi}\left( \frac{x-\mu }{\sigma} \right)$ for a location parameter $\mu \in \mathbb{R}$, a scale parameter $\sigma >0$, and a shape parameter $\xi \in \mathbb{R}$.
\end{Definition}

\subsection{Distribution Of The Exceedances}
Instead of considering only the maximum from a block, the peaks-over-threshold method rather considers all the exceedances over some high threshold value $u$.

\begin{Definition}{\textbf{Excess Distribution over threshold $\boldsymbol u$.}}\\
Let $X$ be a random variable with distribution function $F$ and an upper end-point $x_F \leq \infty$, then the excess distribution over the threshold $u$ is defined as 
\begin{align*}
F_u(x) = \text{Pr}(X-u\leq x|X>u)=\frac{F(x+u)-F(u)}{1-F(u)}, \ \ \  0\leq x< x_F-u.
\end{align*}
\end{Definition}

\begin{Theorem} \autocite{pickands1975statistical,balkema1974residual}\label{Thm:GPD} \\ 
If $F$ is a distribution function that belongs to the maximum domain attraction of a GEV distribution $H_{\xi,\mu,\sigma}$, i.e. if $F$ satistifies the conditions of Theorem \ref{Thm: FisherTippet}, then 
$$\lim_{u\to x_F}\sup_{0\leq x<x_F-u}\left| F_u(x) -G_{\xi,\sigma}(x)  \right|=0,$$
where \begin{align*}
G_{\xi,\sigma}(x)&=
\begin{cases} 
      1-\left(1+\xi\frac{x}{\sigma}  \right)^{-\frac{1}{\xi}},   &\ \   \xi\neq0,\\
      1-\exp\left(-\frac{x}{\sigma}  \right),    &\ \  \xi=0,\\
  \end{cases}
\end{align*}
for $\sigma>0$, and $x\geq0$ when $\xi\geq0$, while $0\leq x\leq -\sigma/\xi$ when $\xi<0$. The parameters $\xi$ and $\sigma$ are referred to, respectively, as the shape and scale parameters.
\end{Theorem}
We assume $F_u \approx G_{\xi,\sigma}$ for a high threshold $u$, where $G_{\xi,\sigma}$ is called the Generalized Pareto Distribution (GPD).

A large quantile is defined as the return level $x_m$ that is exceeded on average once every $m$ observations provided that $m$ is large enough and satisfying $x_m>u$, where
\begin{align*}
  \text{Pr}(X>x_m) = \text{Pr}(X>u) \cdot \left[ 1 + \xi \left( \frac{x_m - u}{\sigma}\right) \right]^{-1/\xi} = \frac{1}{m}.
\end{align*}
By solving for $x_m$,
\begin{equation}\label{Eq: GPDQuantile}
    x_m = u + \frac{\sigma}{\xi} \left[ \left(m \cdot\text{Pr}(X>u)\right)^\xi -1\right].
\end{equation}

\subsection{Poisson Approximation of the Number of Excesses over a High Threshold}
\begin{Theorem}\label{Thm:PoissonGPD}
Let $X_1,\ldots,X_n$ be a sequence of independent random variables satisfying the conditions in Theorem \ref{Thm:GPD}, and let $N_n$ be the number of excesses over a threshold
$u_n$. If the sequence of threshold $(u_n)$ satisfies 
$$\lim_{n\to\infty} n(1-F(u_n)) = \lambda,$$
then, for $k=0,1,2,\ldots,$
$$\lim_{n\to\infty}\P(N_n\leq k) = \sum_{s=0}^k \frac{e^{-\lambda}\lambda^s}{s!}.$$
\end{Theorem}
Thus, assuming a high threshold $u$, the number of exceedances can be modelled as a marked homogeneous Poisson process with rate $\lambda$ and the size of the excesses is the limiting GPD. Accordingly, the point process model requires the estimation of the rate $\lambda$ of the homogeneous Poisson process and the parameters of the GPD.

\subsection{Distribution Of The Maximum Of Homogeneous Poisson Process With GPD Marks} \label{Section: MaxofGPD}

Let $N$ be a Poisson random variable with mean $\lambda$ and let $X_1, \ldots, X_N$ be a sequence of $N$ independent and identically
distributed random variables with common cumulative distribution function $G_{\xi,\sigma}$. Let $M_N = \max\lbrace X_1,\ldots,X_N\rbrace$, then \citet{cebrian2003} indicates that
$$\P(M_N \leq x) = H_{\xi, \mu, \psi}(x),$$
with $\mu = \sigma/\xi\left(\lambda^\xi -1\right)$ and $\psi = \sigma \lambda^\xi$.

\section{Algorithms} \label{Appendix: Algorithms}
\begin{algorithm}[H] \label{Algorithm: EQSimulation}
\SetAlgoVlined
\KwResult{$\PML_{1/\epsilon}$ of Eastern and Western Canada earthquake losses and claims}
\SetKwData{Left}{left}
\SetKwData{This}{this}
\SetKwData{Up}{up}
\SetKwFunction{CalculateLossesAndClaims}{CalculateLossesandClaims}
\nl $n \leftarrow 100000$\;
\nl \textbf{SpatioTemporalSim}$\leftarrow n$ years of simulated earthquakes from a spatio-temporal point process\;
\nl \For{$i\leftarrow 1$ \KwTo $\length(\textbf{SpatioTemporalSim})$}{
\nl $\textbf{SimEQ}_i\leftarrow$ coordinates of the $i^{\textup{th}}$ earthquake in \textbf{SpatioTemporalSim}\;
\nl E$_i\leftarrow \mathbbm{1}_{\left(\textup{Longitude of } \textbf{SimEQ}_i> -100\right)}$\;
\nl $\textbf{Grid}_{i} \leftarrow$ coordinates of the nearest neighbor from the PGA Grid to $\textbf{SimEQ}_{i}$\; 
\nl PGA$_{i}\leftarrow$ a simulated PGA value from a fitted GPD, whose parameters are estimated from the 8 PGA quantiles in \textbf{Grid}$_i$. PGA$_i$ must be corresponding to a moment magnitude $>6$.\;  
\nl MMI$_{i}\leftarrow 3.66 \log(\text{PGA}_i) -1.66,$ for PGA$_i$ in cm/s$^2$\;
\nl d$_{i} \leftarrow$ distance in km between ${\textbf{SimEQ}}_{i}$ and ${\textbf{Grid}}_{i}$.\; 
\nl \eIf{E$_{i} == 1$}{
        \nl $\text{Magnitude}_i = \left(\text{MMI}_i - 1.41 + 2.08\log_{10}(\text{d}_i) + 0.00345 \text{d}_i\right)/1.68$,\;}{
       \nl  $\text{Magnitude}_i = (\text{MMI}_i - 5.07 + 3.69\log_{10}(\text{d}_i))/1.09$,\;}
\nl $\textbf{RadiusMMI}_i\leftarrow$ the radii of MMI circles centered at ${\textbf{SimEQ}}_{i}$ for MMI levels from $\text{MMI}_{i}$ to \rom{6}, for a given value $\text{Magnitude}_i$.\;
\nl \For{$j\leftarrow 1$ \KwTo $\length({\textbf{RadiusMMI}}_{i})$}{
\nl $\textbf{CSD}_{i,j}\leftarrow$ list of CSDs that intersect MMI circle of radius ${\textbf{RadiusMMI}}_{i,j}$\;
\nl ${\textbf{CSDAreaPrcnt}}_{i,j} \leftarrow$ the percentage of land of each CSD that intersects the MMI circle of radius ${\textbf{RadiusMMI}}_{i,j}$\;
\nl \CalculateLossesAndClaims($\text{MMI}_i$-[j-1],${\textbf{CSD}}_{i,j},{\textbf{CSDAreaPrcnt}}_{i,j}$)}
}
\caption{Earthquake losses and claims simulation}\label{EQ_Sim}
\end{algorithm}
\begin{algorithm}[H]
\caption{Calculate Losses and Claims Function}\label{Algorithm: EQLossesClaims}
\SetKwData{MDF}{MDF}
\SetKwData{MMI}{MMI}
\SetKwData{mmi}{mmi}
\SetKwData{CSD}{\textbf{CSD}}
\SetKwData{Loss}{Loss}
\SetKwData{LOSS}{LOSS}
\SetKwData{CLAIM}{CLAIM}
\SetKwData{Ded}{Ded}
\SetKwData{DamageRange}{DamageRange}
\SetKwData{pDamageRange}{pDamageRange}
\SetKwData{BldgCostUncertainty}{BldgCostUncertainty}
\SetKwData{DedPrcnt}{DedPrcnt}
\SetKwData{LmtPrcnt}{LmtPrcnt}
\SetKwData{InsPrcnt}{InsPrcnt}
\SetKwData{Lmt}{Lmt}
\SetKwData{CSDAreaPrcnt}{\textbf{CSDAreaPrcnt}}
\SetKwFunction{CalculateLossesAndClaims}{CalculateLossesAndClaims}
\SetKwInOut{Input}{Input}
\SetKwInOut{Output}{Output}
\Input{MMI of earthquake (MMI), list of CSDs (\textbf{CSD}), percentage affected for each CSD (\textbf{CSDAreaPrcnt})}
\Output{Losses and Claims}
\BlankLine
\SetKwFor{foreach}{foreach}{do}{}
\nl\foreach{CSD $C \in \textbf{CSD}$}{
\nl All steps are done for \textbf{CSD} $C$, but the subscript is removed for readability.\;
\nl Let $\text{Loss}_{t,B}$ be the losses for building type $B$ and damage type $t$, for $t\in\lbrace S, AS, DS, BldgC \rbrace$.\;
\nl Let $\text{Exposure}_{t,B}$ be the building and/or building contents exposure for building type $B$ and damage type $t$, for $t\in\lbrace S, AS, DS, BldgC \rbrace$.\;
\nl Let $\text{DamageRange}_{t,B,i}$ be the damage range associated with damage state $i$, building type $B$, and damage type $t$, for $t\in\lbrace S, AS, DS, BldgC \rbrace$ for a given $MMI$ level. Let $\text{pDamageRange}_{t,B,i}$ be its associated probability.\;
\nl Let $\text{MDF}_{t,B,i}$ be the mean damage factor associated with damage state $i$ for building type $B$ and damage type $t$, for $t\in\lbrace S, AS, DS, BldgC \rbrace$.\;
\nl Let $\text{DedPrcnt}$ be the percentage deductible\;
\nl Let $\text{LmtPrcnt}$ be the percentage policy limit\;
\nl Let $\text{InsPrcnt}$ be the percentage of insurance market penetration\;
\nl \foreach{Building type $B$}{
\nl Let $\text{BldgCostUncertainty}$ be the uncertainty associated to the cost of building replacement.\;
    \nl $\text{BldgCostUncertainty} \leftarrow \mathcal{U}(0.9,1.1)$\;
    \nl $\text{Ded}_{B} \leftarrow \textbf{CSDAreaPrcnt}\times\text{DedPrcnt}\times\sum_t \text{Exposure}_{t,B} \times\text{BldgCostUncertainty} $\;
    \nl $\text{Lmt}_{B} \leftarrow \textbf{CSDAreaPrcnt}\times\text{LmtPrcnt}\times\sum_t \text{Exposure}_{t,B} \times\text{BldgCostUncertainty} $\;
    \nl \foreach{Damage type $t \in \lbrace \text{S, AS, DS, BldgC} \rbrace$}{
    \nl $\text{MDF}_{t,B,i}\leftarrow \left[\text{sample from } \mathcal{U}\left(\text{DamageRange}_{t,B,i}\right)\right]\times \text{pDamageRange}_{t,B,i}$\;
    \nl $\text{MDF}_{t,B}\leftarrow \sum_i \text{MDF}_{t,B,i}$\;
    \nl $\text{Loss}_{t,B} \leftarrow \text{Exposure}_{t,B} \times\text{BldgCostUncertainty} \times \text{MDF}_{t,B} \times \textbf{CSDAreaPrcnt}$\;}
    \nl $\text{LOSS}_B\leftarrow \sum_t \text{Loss}_{t,B}$\;
    \nl     $\text{CLAIM}_B\leftarrow \text{InsPrcnt}\times \max\lbrace0,\min\lbrace\text{LOSS}_B - \text{Ded}_B, \text{Lmt}_B -\text{Ded}_B\rbrace\rbrace$\;
    }
    \nl $\text{LOSS} = \sum_B \text{LOSS}_B$\;
\nl     $\text{CLAIM} = \sum_B \text{CLAIM}_B$\;
}
\nl \KwRet{$\sum_C \text{LOSS}$,$\sum_C \text{CLAIM}$}
\end{algorithm}
\newpage

\section{Correlation of losses and claims} \label{Appendix: CorrelationLossesClaims}

\begin{table}[ht]
\centering\captionof{table}{Pearson correlation coefficient of the simulated insurance claims between Canadian provinces, based on 100,000 years of simulated earthquakes.}
\label{Table: ClaimsCorrelation}
\begin{tabular}{|r|rrrrrrrrrrrrr|}
  \hline
 & \textbf{NL} & \textbf{PE} & \textbf{NS} & \textbf{NB} & \textbf{QC} & \textbf{ON} & \textbf{MB} & \textbf{SK} & \textbf{BC} & \textbf{YT} & \textbf{NT} & \textbf{AB} & \textbf{NU} \\ 
  \hline
\textbf{NL} & 1.00 & 0.29 & 0.23 & 0.23 & \multicolumn{1}{|c}{0.00} & 0.00 & 0.00 & 0.00 & 0.00 & 0.00 & 0.00 & 0.00 & 0.00 \\ 
  \textbf{PE} & 0.29 & 1.00 & 0.77 & 0.89 & \multicolumn{1}{|c}{0.00} & 0.00 & 0.00 & 0.00 & 0.00 & 0.00 & 0.00 & 0.00 & 0.00 \\ 
  \textbf{NS} & 0.23 & 0.77 & 1.00 & 0.87 & \multicolumn{1}{|c}{0.00} & 0.00 & 0.00 & 0.00 & 0.00 & 0.00 & 0.00 & 0.00 & 0.00 \\ 
  \cline{6-6}
  \textbf{NB} & 0.23 & 0.89 & 0.87 & 1.00 & 0.02 & \multicolumn{1}{|c}{0.00} & 0.00 & 0.00 & 0.00 & 0.00 & 0.00 & 0.00 & 0.00 \\ 
  \cline{2-4}
  \cline{7-8}
  \textbf{QC} & 0.00 & 0.00 & \multicolumn{1}{c|}{0.00} & 0.02 & 1.00 & 0.69 & 0.48 & \multicolumn{1}{|c}{0.00} & 0.00 & 0.00 & 0.00 & 0.00 & 0.00 \\ 
  \cline{5-5}
  \textbf{ON} & 0.00 & 0.00 & 0.00 & \multicolumn{1}{c|}{0.00} & 0.69 & 1.00 & 0.60 & \multicolumn{1}{|c}{0.00} & 0.00 & 0.00 & 0.00 & 0.00 & 0.00 \\ 
  \cline{9-14}
  \textbf{MB} & 0.00 & 0.00 & 0.00 & \multicolumn{1}{c|}{0.00} & 0.48 & 0.60 & 1.00 & 0.01 & 0.00 & 0.00 & 0.03 & 0.00 & 0.03 \\ 
  \cline{6-7}
  \textbf{SK} & 0.00 & 0.00 & 0.00 & 0.00 & 0.00 & \multicolumn{1}{c|}{0.00}&  0.01 & 1.00 & 0.01 & 0.02 & 0.09 & 0.01 & 0.07 \\ 
  \textbf{BC} & 0.00 & 0.00 & 0.00 & 0.00 & 0.00 & \multicolumn{1}{c|}{0.00} & 0.00 & 0.01 & 1.00 & 0.16 & 0.08 & 0.63 & 0.11 \\ 
  \textbf{YT} & 0.00 & 0.00 & 0.00 & 0.00 & 0.00 & \multicolumn{1}{c|}{0.00} & 0.00 & 0.02 & 0.16 & 1.00 & 0.39 & 0.06 & 0.34 \\ 
  \textbf{NT} & 0.00 & 0.00 & 0.00 & 0.00 & 0.00 & \multicolumn{1}{c|}{0.00} & 0.03 & 0.09 & 0.08 & 0.39 & 1.00 & 0.04 & 0.82 \\ 
  \textbf{AB} & 0.00 & 0.00 & 0.00 & 0.00 & 0.00 & \multicolumn{1}{c|}{0.00} & 0.00 & 0.01 & 0.63 & 0.06 & 0.04 & 1.00 & 0.07 \\ 
  \textbf{NU} & 0.00 & 0.00 & 0.00 & 0.00 & 0.00 & \multicolumn{1}{c|}{0.00} &  0.03 & 0.07 & 0.11 & 0.34 & 0.82 & 0.07 & 1.00 \\ 
   \hline
\end{tabular}
\end{table}

\begin{table}[!ht]
\centering\captionof{table}{Kendall's tau of the simulated financial losses between Canadian provinces, based on 100,000 years of simulated earthquakes.}
\label{Table: LossesCorrelationKendall}
\begin{tabular}{|r|rrrrrrrrrrrrr|}
 \hline
 & \textbf{NL} & \textbf{PE} & \textbf{NS} & \textbf{NB} & \textbf{QC} & \textbf{ON} & \textbf{MB} & \textbf{SK} & \textbf{BC} & \textbf{YT} & \textbf{NT} & \textbf{AB} & \textbf{NU} \\ 
  \hline
\textbf{NL} & 1.00 & 0.73 & 0.75 & 0.39 & 0.22 & 0.09 & \multicolumn{1}{|c}{0.00} & 0.00 & 0.00 & 0.00 & 0.00 & 0.00 & 0.00 \\ 
  \textbf{PE} & 0.73 & 1.00 & 0.78 & 0.52 & 0.33 & 0.19 & \multicolumn{1}{|c}{0.00} & 0.00 & 0.00 & 0.00 & 0.00 & 0.00 & 0.00 \\ 
  \textbf{NS} & 0.75 & 0.78 & 1.00 & 0.53 & 0.33 & 0.19 & \multicolumn{1}{|c}{0.00} & 0.00 & 0.00 & 0.00 & 0.00 & 0.00 & 0.00 \\ 
  \cline{8-8}
  \textbf{NB} & 0.39 & 0.52 & 0.53 & 1.00 & 0.74 & 0.65 & 0.19 & \multicolumn{1}{|c}{0.00} & 0.00 & 0.00 & 0.00 & 0.00 & 0.00 \\ 
  \textbf{QC} & 0.22 & 0.33 & 0.33 & 0.74 & 1.00 & 0.88 & 0.43 & \multicolumn{1}{|c}{0.00} & 0.00 & 0.00 & 0.00 & 0.00 & 0.00 \\ 
  \cline{9-9}
  \textbf{ON} &0.09 & 0.19 & 0.19 & 0.65 & 0.88 & 1.00 & 0.51 & 0.01 & \multicolumn{1}{|c}{0.00} & 0.00 & 0.00 & 0.00 & 0.00 \\ 
  \cline{2-4}
  \cline{10-14}
  \textbf{MB} & 0.00 & 0.00 & \multicolumn{1}{c|}{0.00} & 0.19 & 0.43 & 0.51 & 1.00 & 0.30 & 0.02 & 0.27 & 0.27 & 0.09 & 0.30 \\ 
  \cline{5-6}
  \textbf{SK} & 0.00 & 0.00 & 0.00 & 0.00 & \multicolumn{1}{c|}{0.00} & 0.01 & 0.30 & 1.00 & 0.20 & 0.57 & 0.61 & 0.44 & 0.43 \\ 
  \cline{7-7}
  \textbf{BC} & 0.00 & 0.00 & 0.00 & 0.00 & 0.00 & \multicolumn{1}{c|}{0.00} & 0.02 & 0.20 & 1.00 & 0.34 & 0.27 & 0.39 & 0.36 \\ 
  \textbf{YT} & 0.00 & 0.00 & 0.00 & 0.00 & 0.00 & \multicolumn{1}{c|}{0.00} & 0.27 & 0.57 & 0.34 & 1.00 & 0.79 & 0.55 & 0.65 \\ 
  \textbf{NT} & 0.00 & 0.00 & 0.00 & 0.00 & 0.00 & \multicolumn{1}{c|}{0.00} & 0.27 & 0.61 & 0.27 & 0.79 & 1.00 & 0.45 & 0.70 \\ 
  \textbf{AB} & 0.00 & 0.00 & 0.00 & 0.00 & 0.00 & \multicolumn{1}{c|}{0.00} & 0.09 & 0.44 & 0.39 & 0.55 & 0.45 & 1.00 & 0.36 \\ 
  \textbf{NU} & 0.00 & 0.00 & 0.00 & 0.00 & 0.00 & \multicolumn{1}{c|}{0.00} & 0.30 & 0.43 & 0.36 & 0.65 & 0.70 & 0.36 & 1.00 \\ 
   \hline
\end{tabular}
\end{table}

\begin{table}[!ht]
\centering\captionof{table}{Kendall's tau of the simulated insurance claims between Canadian provinces, based on 100,000 years of simulated earthquakes.}
\label{Table: ClaimsCorrelationKendall}
\begin{tabular}{|r|rrrrrrrrrrrrr|}
 \hline
 & \textbf{NL} & \textbf{PE} & \textbf{NS} & \textbf{NB} & \textbf{QC} & \textbf{ON} & \textbf{MB} & \textbf{SK} & \textbf{BC} & \textbf{YT} & \textbf{NT} & \textbf{AB} & \textbf{NU} \\ 
  \hline
\textbf{NL} & 1.00 & 0.69 & 0.70 & 0.36 & 0.21 & 0.08 & \multicolumn{1}{|c}{0.00} & 0.00 & 0.00 & 0.00 & 0.00 & 0.00 & 0.00 \\ 
  \textbf{PE} & 0.69 & 1.00 & 0.82 & 0.51 & 0.33 & 0.17 & \multicolumn{1}{|c}{0.00} & 0.00 & 0.00 & 0.00 & 0.00 & 0.00 & 0.00 \\ 
  \textbf{NS} & 0.70 & 0.82 & 1.00 & 0.52 & 0.34 & 0.17 & \multicolumn{1}{|c}{0.00} & 0.00 & 0.00 & 0.00 & 0.00 & 0.00 & 0.00 \\ 
  \cline{8-8}
  \textbf{NB} & 0.36 & 0.51 & 0.52 & 1.00 & 0.75 & 0.65 & 0.20 & \multicolumn{1}{|c}{0.00} & 0.00 & 0.00 & 0.00 & 0.00 & 0.00 \\ 
  \textbf{QC} & 0.21 & 0.33 & 0.34 & 0.75 & 1.00 & 0.87 & 0.46 & \multicolumn{1}{|c}{0.00} & 0.00 & 0.00 & 0.00 & 0.00 & 0.00 \\ 
  \cline{9-9}
  \textbf{ON} & 0.08 & 0.17 & 0.17 & 0.65 & 0.87 & 1.00 & 0.55 & 0.04 & \multicolumn{1}{|c}{0.00} & 0.00 & 0.00 & 0.00 & 0.00 \\ 
  \cline{2-4}
  \cline{10-14}
  \textbf{MB} & 0.00 & 0.00 & \multicolumn{1}{c|}{0.00} & 0.20 & 0.46 & 0.55 & 1.00 & 0.22 & 0.00 & 0.15 & 0.15 & 0.00 & 0.19 \\ 
  \cline{5-6}
  \textbf{SK} & 0.00 & 0.00 & 0.00 & 0.00 & \multicolumn{1}{c|}{0.00} & 0.04 & 0.22 & 1.00 & 0.11 & 0.36 & 0.50 & 0.21 & 0.30 \\ 
  \cline{7-7}
  \textbf{BC} & 0.00 & 0.00 & 0.00 & 0.00 & 0.00 & \multicolumn{1}{c|}{0.00} & 0.00 & 0.11 & 1.00 & 0.27 & 0.21 & 0.33 & 0.30 \\ 
  \textbf{YT} & 0.00 & 0.00 & 0.00 & 0.00 & 0.00 & \multicolumn{1}{c|}{0.00} & 0.15 & 0.36 & 0.27 & 1.00 & 0.71 & 0.34 & 0.57 \\ 
  \textbf{NT} & 0.00 & 0.00 & 0.00 & 0.00 & 0.00 & \multicolumn{1}{c|}{0.00} & 0.15 & 0.50 & 0.21 & 0.71 & 1.00 & 0.23 & 0.59 \\ 
  \textbf{AB} & 0.00 & 0.00 & 0.00 & 0.00 & 0.00 & \multicolumn{1}{c|}{0.00} & 0.00 & 0.21 & 0.33 & 0.34 & 0.23 & 1.00 & 0.23 \\ 
  \textbf{NU} & 0.00 & 0.00 & 0.00 & 0.00 & 0.00 & \multicolumn{1}{c|}{0.00} & 0.19 & 0.30 & 0.30 & 0.57 & 0.59 & 0.23 & 1.00 \\ 
   \hline
\end{tabular}
\end{table}

\end{appendices}

\clearpage
\end{sloppypar}
\end{document}